\documentclass[reprint,twocolumn,superscriptaddress,amsmath,amssymb,prb,nofootinbib]{revtex4-1}

\usepackage[dvipdfmx]{graphicx}
\usepackage{dcolumn}
\usepackage{bm}
\usepackage{color}
\usepackage{amssymb}
\usepackage{comment}
\usepackage{ulem}

\begin{document}

\title{Saturable absorption in highly excited silicon and its suppression at the surface}

\author{Shunsuke Yamada}
\affiliation{Kansai Institute for Photon Science, National Institutes for Quantum Science
and Technology (QST), Kyoto 619-0215, Japan}
\author{Tomohito Otobe}
\affiliation{Kansai Institute for Photon Science, National Institutes for Quantum Science
and Technology (QST), Kyoto 619-0215, Japan}

\date{\today}

\begin{abstract}
Nonlinear electronic excitation in laser-irradiated silicon at finite electron temperatures is numerically investigated by first-principles calculations based on the time-dependent density functional theory.
In bulk silicon at finite temperatures under near-infrared laser irradiation, we found that the absorbed energy is saturated when using a certain laser intensity even with a few-cycle pulse. 
Although one-photon processes of conduction-to-conduction and valence-to-valence transitions
are dominant at such a laser intensity, the Pauli blocking inhibits further one-photon transition. 
With higher intensities, multi-photon excitation across the bandgap overwhelms the one-photon excitation and the saturable absorption disappears.
At the surface of finite-temperature silicon, the Pauli blocking is suppressed by the symmetry breaking and the absorbed energy is relatively enhanced from the energy of the saturable absorption in the bulk region.
\end{abstract}
\maketitle


\section{Introduction}

Material properties under laser irradiation have been attracting considerable interest from both fundamental and applied points of view.
Especially, laser ablation encompasses a broad range of physics, including optics, quantum mechanics, material science, and thermodynamics.
Femtosecond laser ablation\cite{Tien1999,Bonse2002,Gattass2008} is essential for precise material processing.
Laser ablation techniques with multiple femtosecond laser pulses\cite{Kerse2016} have been widely used and those improvements are intensively investigated\cite{Obata2023}.

In a laser irradiation process of a material with a femtosecond pulse train, the electron thermalization caused by preceding pulses should have a significant impact on the nonlinear light absorption of subsequent pulses.
For example, a laser ablation process with femtosecond laser pulses is initiated by the energy transfer from the electromagnetic field to the electron system in the femtosecond time scale.
Excited electrons and holes are thermalized at the next time scale and the lattice will be thermalized at even later times.
Because the time scales of thermal equilibrium for electrons and phonons are separated\cite{Goldman1994,Mueller2013,Ramer2014,Rethfeld2017}, there is a time when transient properties of the nonlinear light absorption at finite electron temperatures but at zero lattice temperature have a fundamental importance.
If there is an unexpected change in such transient properties, it may affect the efficiency of the process when the time interval of pulses is in that time scale. 
In that sense, there may be cumulative effects in a laser ablation process that have not been taken into account in existing theoretical methods such as the two-temperature model\cite{Rethfeld2017}. 
Although there is a first-principles computational study of the linear response in a laser-excited semiconductor at finite electron temperatures\cite{Sato2014}, the nonlinear light absorption in finite-temperature systems has not been systematically studied.

In this paper, we numerically investigate the nonlinear electronic excitation in laser-irradiated silicon (Si) at finite electron temperatures.
We employ first-principles computational methods based on the time-dependent density functional theory (TDDFT)\cite{Runge1984} for calculating the real-time electronic motion in Si under the presence of a femtosecond laser pulse.
In TDDFT calculations of the present work, the electron temperature is taken into account via the Fermi-Dirac distribution and the atomic motion is ignored, where atoms are fixed at their equilibrium positions at zero temperature.
Firstly, we discuss how the electron temperature affects linear and nonlinear light absorption in the bulk unit cell of Si.
The physical mechanisms behind its properties are clarified by qualitatively reproducing the TDDFT results with a simplified toy model.

To clarify the correspondence between TDDFT calculations and experiments, it is critical to consider the effects of the surface electronic structure and nonlinear light propagation on light absorption
We evaluate the surface effect by comparing calculation results between the cases of bulk Si and slab Si systems.
We also discuss the light propagation effect by using combined methods of TDDFT for electrons and the Maxwell equations for the electromagnetic field\cite{Yabana2012,Yamada2018}.
As a preparation for future works, we assess the applicability of the combined methods to the nonlinear light propagation in finite-temperature systems by calculating nanoscale Si slabs.

This paper is organized as follows: Sec.~\ref{sec:method} describes
theoretical methods and numerical details.  
In Sec.~\ref{sec:results1} and \ref{sec:results2},
the calculation results are presented and analyzed in detail.
Finally, a conclusion is presented in Sec.~\ref{sec:conclusion}.

\section{Theoretical methods \label{sec:method}}

\subsection{TDDFT in bulk and slab systems \label{sec:method_tddft}}

\subsubsection{Bulk unit cell\label{sec:method_tddft_bulk}}

We first consider the electronic motion in a unit cell of
a crystalline solid driven by a pulsed electric field ${\bf E}(t)$ with a
given time profile.  
We assume that ${\bf E}(t)$ is spatially uniform (dipole approximation) and atoms in the unit cell are fixed at their equilibrium positions at zero temperature.
In the real-time TDDFT, the electronic motion
is described by the following time-dependent Kohn-Sham (TDKS) equation
for the time-dependent Bloch orbitals $u_{n{\bf k}}({\bf r},t)$
\cite{Bertsch2000,Otobe2008}:
\begin{eqnarray}
\begin{aligned}
    i\hbar \frac{\partial}{\partial t} u_{n{\bf k}}({\bf r},t)
        &= \hat{H}^{\rm KS}_{\bf k}(t) u_{n{\bf k}}({\bf r},t), \\
    \hat{H}^{\rm KS}_{\bf k}(t)    
        &= \frac{1}{2m_{\rm e}} \left( -i\hbar \nabla + \hbar{\bf k} + \frac{e}{c} {\bf A}(t) \right)^2  \\    
    & \quad - e\phi({\bf r},t)
        + \hat v_{\rm{NL}}^{{{\bf k}+\frac{e}{\hbar c}{\bf A}(t)}} 
        + V_{\rm{xc}}({\bf r},t), 
\label{eq:tdks}
\end{aligned}
\end{eqnarray}
where $n$ and ${\bf k}$ are the band index and $k$-vector, respectively.
${\bf A}(t)$ is the vector potential which is related to the electric field as ${\bf E}(t)=-(1/c)d{\bf A}(t)/dt$.
Here we treat the dynamics of the valence electrons with the norm-conserving
pseudopotential \cite{Troullier1991}.  
The scalar potential $\phi({\bf r},t)$ consists of the Hartree potential and the local part of the ionic pseudopotential.
The nonlocal part of the pseudopotential is modified as $\hat{v}_{{\rm NL}}^{{\bf k}}\equiv e^{-i{\bf k}\cdot{\bf r}}\hat{v}_{{\rm NL}}e^{i{\bf k}\cdot{\bf r}}$, where $\hat v_{\rm NL}$ is the separable form of the norm-conserving pseudopotential \cite{Kleinman1982}. 
$V_{\rm xc}({\bf r},t)$ is the exchange-correlation potential for which the adiabatic 
local-density approximation (adiabatic LDA) \cite{Perdew1981} is assumed.

The scalar potential $\phi({\bf r},t)$ satisfies the Poisson equation,
\begin{equation}
    \nabla^2 \phi({\bf r},t) = -4\pi (\rho_{\rm ion}({\bf r}) - e n_{\rm
      e}({\bf r},t) ) \ ,
      \label{eq:poisson}
\end{equation}
where ionic charge density $\rho_{\rm ion}({\bf r})$ is prepared so that the local part of the pseudopotential is produced in $\phi$. 
The electron number density $n_{\rm e}({\bf r},t)$ is given by
\begin{equation}
    n_{\rm e}({\bf r},t) = \frac{1}{N_k}\sum_{{\bf k},n}
    f_{n{\bf k}} \vert u_{n{\bf k}}({\bf r},t) \vert^2,
\end{equation}
where $N_k$ and $f_{n{\bf k}}$ are the number of $k$-points and the occupancy at the ground state, respectively.
At the electron temperature $T_e$, $f_{n{\bf k}}$ is given by the following Fermi-Dirac distribution function:
\begin{equation}
    f_{n{\bf k}}=\frac{1}{1+e^{(\varepsilon^{\rm GS}_{n{\bf k}}-\mu)/k_{\rm B}T_e}},
    \label{eq:fermi_dirac}
\end{equation}
where $\varepsilon^{\rm GS}_{n{\bf k}}$ and $\mu$ are the orbital energy at the ground state and chemical potential, respectively.
Throughout this paper, we use the unit of $k_{\rm B}=1$ and express $T_e$ in the unit of energy.

The Bloch orbitals are initially set to the ground state solution: $u_{n{\bf k}}({\bf r},t=0)=u^{\rm GS}_{n{\bf k}}({\bf r})$.
Solving the TDKS equation, we obtain the electric current in the unit-cell volume $\Omega$ as
\begin{equation}
{\bf J}(t)=\int_{\Omega}{d^3 r}\,{\bf j}({\bf r},t) + \Delta{\bf J}_{\rm NL}(t),
\label{eq:J}
\end{equation}
where ${\bf j}({\bf r},t)$ is the microscopic current density,
\begin{eqnarray}
    {\bf j}({\bf r},t) &=& - \frac{e}{m_{\rm e} N_k} {\rm Re }\sum_{{\bf k},n} f_{n{\bf k}} \, u_{n{\bf k}}^*({\bf r},t) \nonumber\\
  && \times  \left\{
    -i\hbar \nabla +\hbar {\bf k} + \frac{e}{c}{\bf A}(t)
\right\} u_{n{\bf k}} ({\bf r},t),
\label{eq:j_micro}
\end{eqnarray}
and $\Delta{\bf J}_{\rm NL}(t)$ is a correction term stemming from the nonlocal potential:
\begin{eqnarray}
\Delta{\bf J}_{\rm NL}(t)&=& -{e}\int_{\Omega} {d^3 r} \frac{1}{N_k}\sum_{{\bf k},n} f_{n{\bf k}} \, u_{n{\bf k}}^*({\bf r},t)  \nonumber\\
&& \times \frac{1}{i\hbar}\left[{\bf r},\hat{v}_{{\rm NL}}^{{{\bf k}+\frac{e}{\hbar c}{\bf A}(t)}}\right ]u_{n{\bf k}}({\bf r},t).
\end{eqnarray}

The total energy of electrons in the unit cell is written as follows:
\begin{eqnarray}
    E_{\rm tot}(t) &=& \int_{\Omega} {d^3 r} \frac{1}{N_k}\sum_{{\bf k},n} f_{n{\bf k}} \, u_{n{\bf k}}^*({\bf r},t) 
    \hat{H}^{\rm KS}_{\bf k}(t) u_{n{\bf k}}({\bf r},t)
    \nonumber \\
    && 
    -\frac{e^2}{2}  \int_{\Omega} {d^3 r} \int_{\Omega}{d^3 r'} \frac{n_{\rm  e}({\bf r},t) n_{\rm e}({\bf r'},t)}{|{\bf r}-{\bf r'}|}  \nonumber \\
    && -  \int_{\Omega} {d^3 r}\,  V_{\rm{xc}}({\bf r},t) n_{\rm  e}({\bf r},t)
    + E_{\rm xc}[n_{\rm  e}](t),
    \label{eq:total_energy}
\end{eqnarray}
where $E_{\rm xc}[n_{\rm  e}](t)$ is the exchange-correlation energy functional.
The second and third terms compensate for the double counting of the Hartree and exchange-correlation potential terms, respectively, in the first term. 
According to the energy conservation, the excitation energy of electrons in the unit cell, $E_{\rm ex}(t)\equiv E_{\rm tot}(t)-E_{\rm tot}(t=0)$, is equal to the absorbed energy $E_{\rm abs}(t)$, or work done by the electric field, as follows:
\begin{equation}
    E_{\rm ex}(t) = E_{\rm abs}(t) \equiv \int^t_0 dt' \,{\bf J}(t')\cdot {\bf E}(t').
\end{equation}

For Si atoms, the correction term $\Delta{\bf J}_{\rm NL}(t)$ is negligible\cite{Yamada2018,Yamada2021,Sato2023}.
For such a case, ignoring $\Delta{\bf J}_{\rm NL}(t)$, we can define the absorbed energy density as follows:
\begin{equation}
    e_{\rm abs}({\bf r},t) \equiv \int^t_0 dt' \,{\bf j}({\bf r},t')\cdot {\bf E}(t').
    \label{eq:eabs}
\end{equation}
This quantity represents the density distribution of the energy transfer from the electromagnetic field to the electron system.

The occupancy of the time-dependent wavefunctions at the time $t$ can be estimated by projecting $\{u_{n{\bf k}}(t)\}$ onto the ground-state orbitals $\{u^{\rm GS}_{n{\bf k}}\}$ as follows:

\begin{equation}
    F_{n{\bf k}}(t) =  \sum_{m} f_{m{\bf k}} |\langle u^{\rm GS}_{n,{\bf k}+\frac{e}{\hbar c}{\bf A}(t)} | u_{m{\bf k}}(t) \rangle|^2,
    \label{eq:occ}
\end{equation}
where we used the Houston function $u^{\rm GS}_{n,{\bf k}+\frac{e}{\hbar c}{\bf A}(t)}$ to suppress the contribution from the virtual excitation.
At $t=0$, $F_{n{\bf k}}(t)$ is equal to the ground state occupancy $f_{n{\bf k}}$, which represents the occupancy by thermal excitation.
To visualize the occupancy change, we define the following weighted density of states:
\begin{equation}
    \tilde{D}(\varepsilon,t) = \frac{1}{N_k}\sum_{{\bf k},n} \Delta F_{n{\bf k}}(t)\delta(\varepsilon-\varepsilon^{\rm GS}_{n{\bf k}}),
    \label{eq:pdos}
\end{equation}
where $\Delta F_{n{\bf k}}(t)=F_{n{\bf k}}(t)-f_{n{\bf k}}$.
Without $\Delta F_{n{\bf k}}(t)$, the above definition is equivalent to the ordinary density of states $D(\varepsilon)$.

\subsubsection{Slab system\label{sec:method_tddft_slab}}

To take into account the effects of the surface electronic structure on the light-matter interaction, we consider a TDDFT computation for a slab system with a thickness of $d$.
We consider a calculation box including the slab sandwiched between vacuum regions and assume the periodic boundary conditions (slab approximation).
The slab surfaces are parallel to the $xy$ plane and the slab material is periodic in the $xy$ plane.
The polarization direction of the applied field ${\bf E}(t)$ is set to the $x$ axis (horizontal axis).

The formalism of the slab TDDFT computation is the same as before (Sec.~\ref{sec:method_tddft_bulk}) except that it uses the calculation box of the slab instead of the unit cell.
The Bloch orbital $u_{n{\bf k}}({\bf r},t)$ is defined in the entire region of the calculation box, where the ${\bf k}$-vector is in the two-dimensional (2D) Brillouin zone of the slab system.
The applied field ${\bf E}(t)$ is assumed to be spatially uniform in the calculation box.
The surface effect can be evaluated by comparing results for the slab with corresponding bulk results.

\subsection{Maxwell-TDDFT methods for light propagation\label{sec:method_light_propagation}}

In the ordinary TDDFT formalism described above, the applied field ${\bf E}(t)$ has a given time profile and the light propagation is not considered. 
If the slab thickness $d$ is much smaller than the laser wavelength, the light propagation effect is negligible.
For a substantial thickness, however, the light propagation strongly modulates the characteristics of the nonlinear light absorption in the material.
We should evaluate the effect of nonlinear light propagation in order to clarify the correspondence of TDDFT results to experimental data.

For considering the nonlinear light propagation in the slab system, it is required to use combined methods of the Maxwell equations for the light propagation dynamics and TDDFT for the electronic motion.
In this paper, we use two such methods, the multiscale\cite{Yabana2012} and single-scale\cite{Yamada2018} Maxwell-TDDFT methods, with and without a coarse-graining approximation, respectively.
The computational cost of the former method is significantly smaller than that of the latter method as a trade-off for the coarse-graining ignoring microscopic details such as the surface effect.
We will assess the applicability of the two methods to finite-temperature systems.
Moreover, we will discuss how to interpret ordinary TDDFT results for predicting experimental measurements (Sec.~\ref{sec:light_propagation}).

Here, we consider laser irradiation of the slab (free-standing thin film in the vacuum) at the normal incidence. 
We assume that the incident laser pulse is a linearly polarized plane wave, where the direction of the light propagation is along the $z$ axis and the polarization direction is along the $x$ axis.

\subsubsection{Multiscale Maxwell-TDDFT method}

The multiscale Maxwell-TDDFT method\cite{Yabana2012} is used to describe the macroscopic light propagation.  
The spatial scale of the light propagation ($\sim$ wavelength) is a few hundred nanometers and is much larger than that of the microscopic electron dynamics.
As in ordinary macroscopic electromagnetism, we apply coarse-graining to the microscopic electronic system.
Ignoring the surface effect, we assume that the electronic motion is local in the bulk region.
In this method, we combine the one-dimensional (1D) wave equation for the light propagation along the $z$ axis and a number of the TDKS equations for the local electronic motion at each point.
The former equation is described as follows:
\begin{equation}
 \left (\frac{1}{c^2} \frac{\partial^2}{\partial t^2}  - \frac{\partial^2}{\partial Z^2} \right) {\bf A}_{Z}(t)
= \frac{4\pi}{c}   {\bf j}_{Z}(t) \ ,
\label{eq:multiscale}
\end{equation}
where ${\bf A}_{Z}(t)$ and ${\bf j}_{Z}(t)$ are the vector potential and local current density, respectively, at the macroscopic coordinate $Z$.
This wave equation is solved by the finite-difference method on a 1D grid of $Z$.
At each grid point $Z$ inside the slab, the local current density ${\bf j}_{Z}(t)$ is calculated by using the TDDFT formalism for the bulk unit cell [Sec.~\ref{sec:method_tddft_bulk}].
Here, ${\bf A}(t)$ in the TDKS equation [Eq.~(\ref{eq:tdks})] is replaced by ${\bf A}_Z(t)$, which is treated as a spatially uniform field in the TDKS system.
${\bf j}_{Z}(t)$ is equal to the current density averaged over the unit cell, ${\bf J}(t)/\Omega$, where ${\bf J}(t)$ is defined in Eq.~(\ref{eq:J}) but replacing ${\bf A}(t)$ with ${\bf A}_Z(t)$.
Eq.~(\ref{eq:multiscale}) and replicas of Eq.~(\ref{eq:tdks}) with ${\bf A}_Z(t)$ are coupled with each other and solved simultaneously at each time step, where the number of replicas is equal to that of the 1D grid points $\{Z\}$ inside the slab.
This method can describe the macroscopic dynamics of the light propagation but the microscopic surface effect is ignored.

\subsubsection{Single-scale Maxwell-TDDFT method}

To consider the microscopic light propagation, we use the single-scale Maxwell-TDDFT method\cite{Yamada2018}.
This method also combines the Maxwell equations and TDDFT but without the coarse-graining.
We consider the same slab system as in Sec.~\ref{sec:method_tddft_slab} and the microscopic vector potential ${\bf A}({\bf r},t)$ satisfying the following wave equation:
\begin{equation}
    \left( \frac{1}{c^2} \frac{\partial^2}{\partial t^2} - \nabla^2 \right) {\bf A}({\bf r},t)
    + \frac{1}{c} \frac{\partial}{\partial t} \nabla \phi({\bf r},t) =  \frac{4\pi}{c} {\bf j}({\bf r},t).
    \label{eq:maxwell_micro}
\end{equation}
The scalar potential $\phi({\bf r},t)$ satisfies the Poisson equation [Eq.~(\ref{eq:poisson})] and these two equations are the microscopic Maxwell equations in the Coulomb gauge.
Here, the microscopic current density ${\bf j}({\bf r},t)$ is defined as Eq.~(\ref{eq:j_micro}) but replacing ${\bf A}(t)$ with ${\bf A}({\bf r},t)$.
The Bloch orbital $ u_{n{\bf k}}({\bf r},t)$ is defined in the calculation box for the slab system and satisfies the TDKS equation [Eq.~(\ref{eq:tdks})], where ${\bf A}(t)$ is replaced by ${\bf A}({\bf r},t)$.
The Maxwell equations and Eq.~(\ref{eq:tdks}) with ${\bf A}({\bf r},t)$ are coupled with each other and solved simultaneously at each time step.
Unlike the multiscale method, the vector potential ${\bf A}({\bf r},t)$ is defined as a microscopic field in the calculation box.
We use the finite-difference method with a three-dimensional (3D) grid of ${\bf r}$ for describing $ u_{n{\bf k}}({\bf r},t)$, ${\bf A}({\bf r},t)$, and $\phi({\bf r},t)$ at each time step $t$.
The propagation of ${\bf A}({\bf r},t)$ in the vacuum far from the surface is described analytically by using the absorbing boundary conditions of the electromagnetic field at the ends of the calculation box in the $z$ axis.
Here, the periodic boundary conditions in the $xy$ plane are imposed on ${\bf A}({\bf r},t)$.
This method can take into account both the effects of the microscopic light propagation and surface electronic structure.

\subsection{Numerical details for TDDFT calculations}

The computational methods described above are implemented in our open-source software package SALMON (Scalable Ab initio
Light-Matter simulator for Optics and
Nanoscience)\cite{Noda2019,SALMON_web} and we use it for numerical simulations below.  
In this code, the TDKS equation is solved by
using the finite-difference method for time and spatial domain.  
The time evolution of the
electron orbitals is carried out using the Taylor expansion method\cite{Yabana1996}.

For TDDFT calculations in bulk Si, we use a cubic unit cell of the diamond structure containing eight Si atoms with the lattice constant of $a = 0.543$ nm.
The numbers of grid points for discretizing the unit cell volume and the Brillouin zone are set to $N_r=16^3$ and $N_k=16^3$, respectively.
The time step is set to $2.5 \times 10^{-3}$ fs.
At the electron temperature $T_e=1$ eV, the number of orbitals per Si atom is set to 6 (2 for occupied and 4 for unoccupied) to obtain converged results.

For TDDFT calculations in a Si slab with the thickness of $d=n_{\rm cell}a$ ($n_{\rm cell}$ is an integer), we use a periodic calculation box with the size of $a\times a\times(d+L_{\rm vac})$, where $L_{\rm vac}$ is the length of the vacuum region.
The cubic unit cell of Si is used as a building block to construct the slab.
The Si slab consists of the $n_{\rm cell}$ blocks aligned along the $z$ axis and has the (001) surfaces.
The atomic positions of Si atoms are set at those positions in the bulk crystalline system,
and the dangling bonds at each (001) surface are terminated by four hydrogen atoms [see Fig. 2(a) in Ref.~\onlinecite{Yamada2018}]. 
Thus, the number of atoms in the calculation box is equal to $8n_{\rm cell}+8$.
We use $L_{\rm vac}=16a$ and the same number of orbitals per Si atom as in the bulk case.
A uniform 3D spatial grid is used with the same grid spacing as that for the bulk unit cell. 
The 2D Brillouin zone of the slab is sampled by a $16\times 16$ $k$-point grid. 

For the single-scale Maxwell-TDDFT method, we use the same calculation conditions as in the slab TDDFT case. 
For the multiscale Maxwell-TDDFT method, a uniform 1D grid is introduced to describe the wave equation Eq.~(\ref{eq:multiscale}), where the number of grid points in the material is set to 16 for the case of $d=30a$.
The parameters for TDKS equations are the same as the bulk Si case.
We have carefully examined the convergence of the calculations with respect to the above parameters.

We employ a pulsed electric field of linear polarization given by the following time profile for the vector potential,
\begin{eqnarray}
{\bf A}(t) = - \frac{c E_0}{\omega} \, \sin \left[ \omega \left( t
  - \frac{\tau_p}{2} \right) \right] \,\, \sin^2 \left( \frac{\pi t}{\tau_p}
\right) \hat{\bf x} , \nonumber \\ (0<t<\tau_p),
\label{eq:pulse}
\end{eqnarray}
where $E_0$, $\omega$, and $\tau_p$ are the peak amplitude of the electric field, central frequency, and pulse duration, respectively.
In TDDFT calculations, ${\bf A}(t)$ defined in Eq.~(\ref{eq:pulse}) is used as the applied field in Eq.~(\ref{eq:tdks}).
For Maxwell-TDDFT calculations, the incident pulse in the vacuum at the front of the slab surface is prepared by Eq.~(\ref{eq:pulse}).

\subsection{Toy model for FTSA\label{sec:method_toy}}

As a simple toy model, we utilize the 1D tight-binding model.
The 1D crystal consists of $N$ periodic cells with the lattice constant $a$. 
Here we assume the periodic boundary condition of the whole system: $f(x+Na) = f(x)$, where $f(x)$ is an arbitrary function defined in the 1D crystal.
We consider localized atomic orbitals $\phi_{b}(x)$ $(b=1,\cdots, N_{\rm band})$ as a basis set for the unit cell at the lattice point $x=0$. 
The wavefunctions of the crystal can be represented as
\begin{equation}
    |\psi_k\rangle=\frac{1}{\sqrt{N}}\sum_{m=0}^{N-1}
    \sum_{b=1}^{N_{\rm band}}
    e^{ikR_m}
|\phi_{b,m}\rangle C_{bk},
\end{equation}
where
$R_m\equiv am$ $(m=0,\cdots,N-1)$ is the position of the lattice point and
$\phi_{b,m}(x) \equiv \phi_{b}(x-R_m)$ is the translated orbital.
The $N$ independent $k$ points are located in the Brillouin zone $[-\pi/a,\pi/a)$.

For each $k$ point, the $N_{\rm band}\times N_{\rm band}$ Hamiltonian matrix is described as
\begin{equation}
    H_{b,b'}[k]=\sum_{m}e^{ikR_m}
\langle \phi_{b,0}|H|\phi_{b',m}\rangle.
\end{equation}
We define the Hamiltonian matrix elements as follows:
\begin{equation}
    \langle \phi_{b,0}|H|\phi_{b',m}\rangle=
  \begin{cases}
    \varepsilon_b, & (m=0,\quad b=b') \\
    -\tilde{t},             & (|m| = 1,\quad b\ne b') \\
 0, & ({\rm otherwise}),
  \end{cases}
\end{equation}
where $\varepsilon_b$ and $\tilde{t}$  are the on-site energy and hopping parameter, respectively.

The electron dynamics under the vector potential $A(t)$ is governed by the following time-dependent Schroedinger equation:
\begin{equation}
    i\hbar \frac{d}{dt}\psi_{bk}(t) = H\left[{k}+\frac{e}{\hbar c}{A}(t)\right] \psi_{bk}(t),
    \label{eq:td_model}
\end{equation}
where $\psi_{bk}(t)$ is a $N_{\rm band}$-dimensional column vector and its initial value is set to the ground state wavefunction: $\psi_{bk}(t=0)=\psi^{\rm GS}_{bk}$.
The number of excited electrons is defined as 
\begin{equation}
    n_{\rm ex}(t) \equiv \frac{1}{N}\sum_k\sum_{b,b'} 
f_{bk} (1-f_{b'k})
|\langle \psi^{\rm GS}_{b'k} | 
\psi_{bk}(t)\rangle |^2,
\label{eq:nex_model}
\end{equation}
where $f_{bk}$ is the Fermi-Dirac distribution function for the eigenstate $\psi^{\rm GS}_{bk}$.

\section{Electronic excitation in the bulk region\label{sec:results1}}

\subsection{TDDFT results for Si bulk \label{sec:tddft_bulk}}

\begin{figure}
    \includegraphics[keepaspectratio,width=\columnwidth]
    {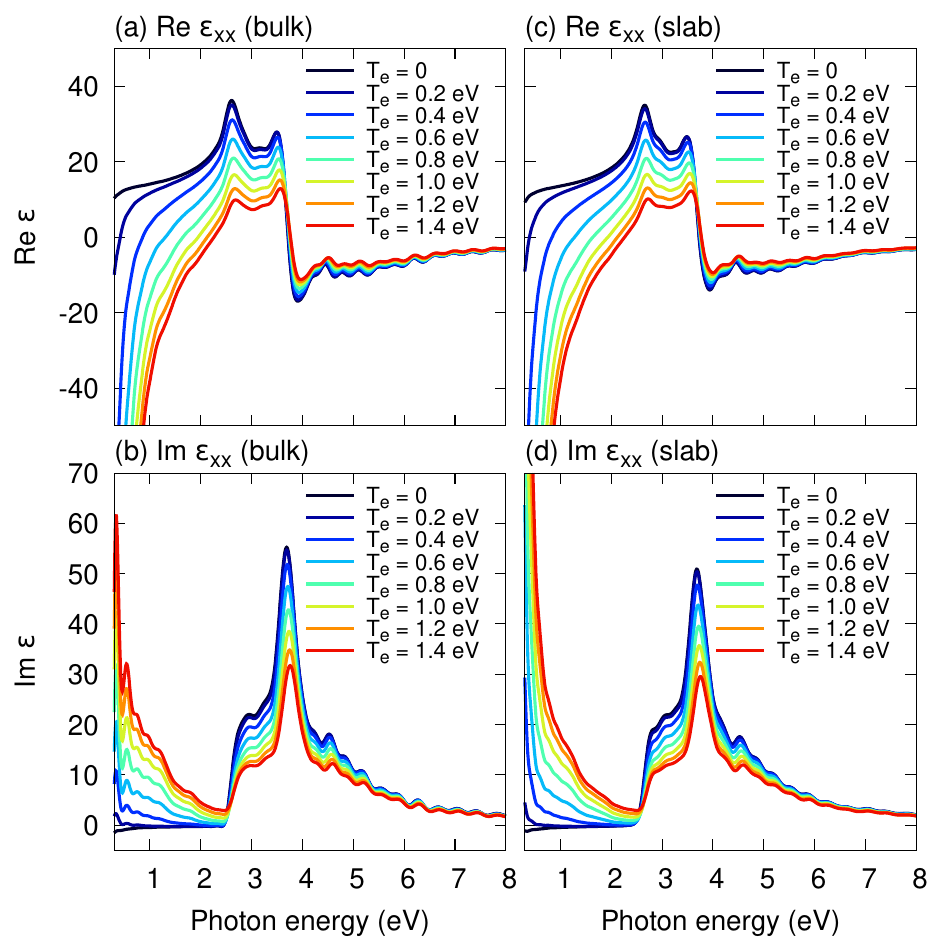}
    \caption{\label{fig:lr}
    (a) Real and (b) imaginary parts of the dielectric function in bulk Si at several electron temperatures.
    (c,d) The same as (a,b) but for the slab Si system.
    }
\end{figure}

First, we focus on the electronic excitation in the bulk region of the laser-irradiated Si.
The dielectric functions of bulk Si at several electron temperatures $T_e$ from $0$ to $1.4$ eV are depicted in Fig.~\ref{fig:lr}(a,b).
In panel (a) [(b)], the real (imaginary) part of the dielectric function decreases (increases) monotonically as the temperature increases at low frequencies.
This originates from the Drude-like response of the thermally excited carriers\cite{Sato2014}.
At $T_e=0$, the bottom edge of the imaginary part corresponds to the optical bandgap of about 2.4 eV, which is underestimated by LDA\cite{Yabana2012} from the experimental value of 3.1 eV.
By thermal excitation, the imaginary part below the bandgap at finite temperatures obtains a nonzero value, which yields one-photon absorption when a light pulse with a below-gap frequency is applied.

\begin{figure}
    \includegraphics[keepaspectratio,width=\columnwidth]
    {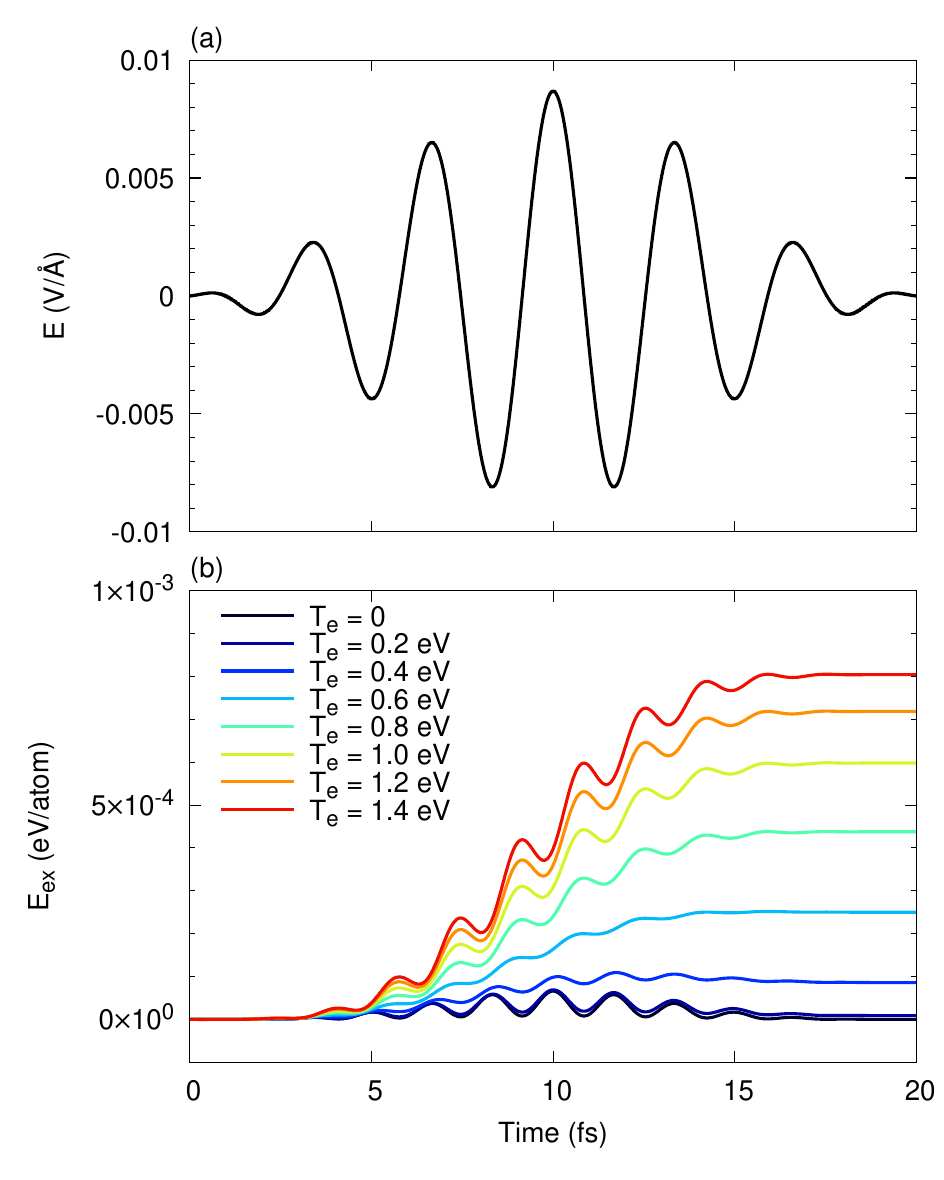}
    \caption{\label{fig:pulse}
    (a) Time profile of the applied electric field. 
    (b) Electronic excitation energy in bulk Si at several temperatures.
    }
\end{figure}

Next, we consider light pulse irradiation.
Figure~\ref{fig:pulse}(a) shows the time profile of the applied electric field with the pulse envelope given by Eq.~(\ref{eq:pulse}), where the pulse duration and photon energy corresponding to the central frequency are set to $\tau_p=20$ fs and $\hbar \omega=1.2$ eV, respectively.
As mentioned above, LDA underestimates the bandgap.
The calculated and experimental values are about 2.4 eV and 3.1 eV, respectively\cite{Yabana2012}.
The influence of the underestimation on physical discussions can be mitigated by scaling down the photon energy of the applied field with the same ratio.
Here, the photon energy $\hbar \omega=1.2$ eV corresponds to 1.55 eV (wavelength 800 nm) as $1.2 = 1.55 \times(2.4/3.1)$.
Additionally, we will see that the qualitative behavior of the electronic excitation at finite temperatures does not depend much on the photon energy of the applied field.
The field amplitude of the electric field, $E_0$, in Fig.~\ref{fig:pulse}(a) is set to provide $I_m=10^9$ W/cm$^2$, where $I_m=cE_0^2/8\pi$ is the peak intensity inside the medium.
Note that $I_m$ differs from the peak intensity of the incident pulse in vacuum, $I_v$, due to the induced electric field in the medium.

Figure~\ref{fig:pulse}(b) shows the electronic excitation energy $E_{\rm ex}(t)$ by the applied field as a function of the time at several electron temperatures.
The excitation energy increases monotonically as the temperature increases.
Because the peak intensity $I_m=10^9$ W/cm$^2$ is sufficiently weak, the behavior of the excitation energy purely reflects the linear response depicted in Fig.~\ref{fig:lr}(a,b).
Even with the below-gap frequency, electrons are substantially excited at finite temperatures.
This is because the broadened occupancy at finite temperatures [Eq.~(\ref{eq:fermi_dirac})] allows one-photon transition inside the conduction band (conduction-to-conduction transition) or valence band (valence-to-valence transition), where partially occupied bands have the space to accept electrons or holes.
This linear one-photon transition is prohibited at zero temperature.
In the linear regime, multi-photon excitation across the bandgap is absent.

\begin{figure*}
    \includegraphics[keepaspectratio,width=\textwidth]{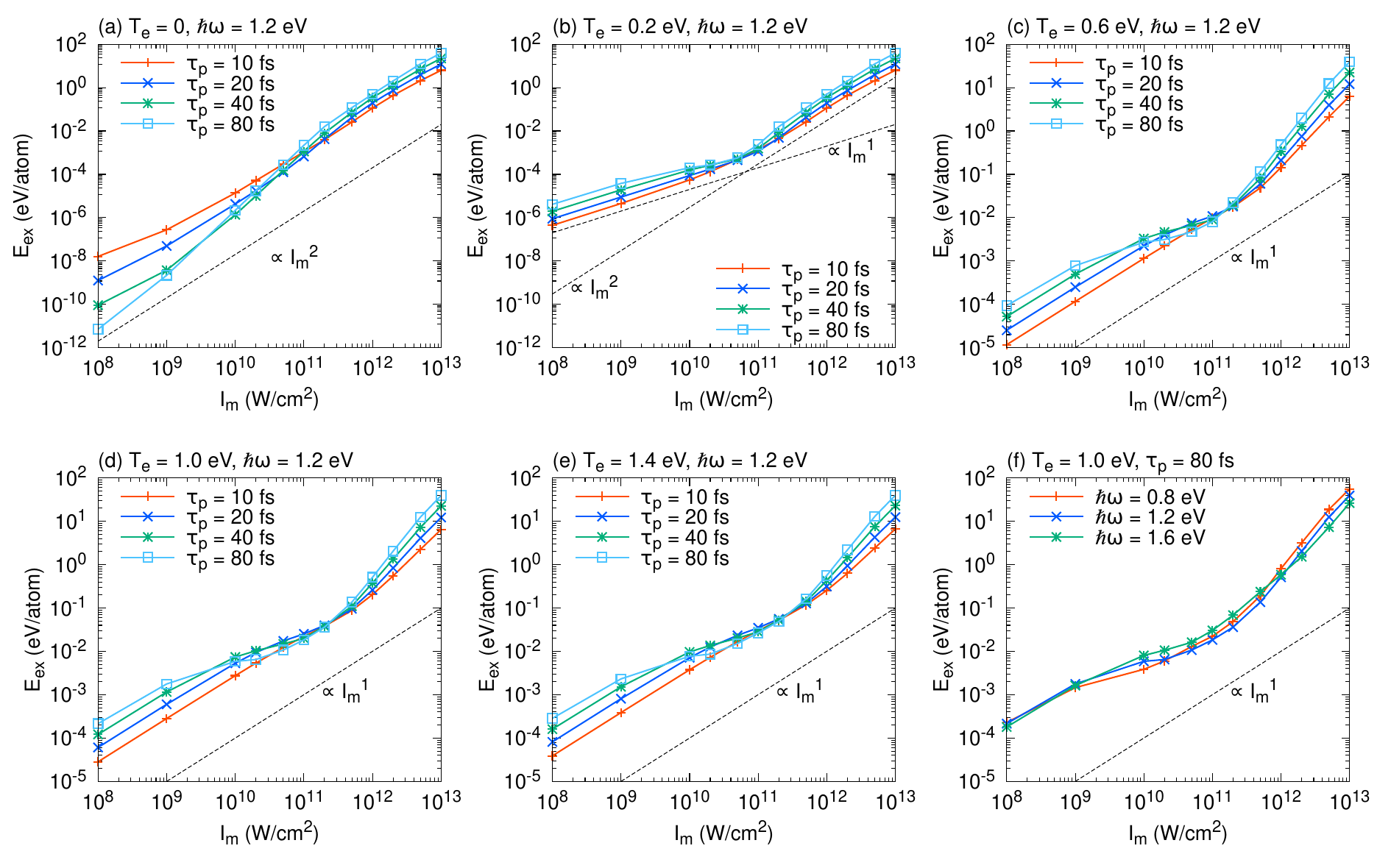}
    \caption{\label{fig:bulk_intensity}
    Logarithmic plots of the excitation energy as a function of the laser intensity in the medium.
    (a--e) The cases for the electron temperatures of $T_e=0$, 0.2, 0.6, 1.0, and 1.4 eV, respectively, with several values of the pulse duration $\tau_p$.
    The photon energy of the applied pulse is set to $\hbar\omega=1.2$ eV.
    (f) $T_e=1$ eV and $\tau_p=80$ fs case but with several photon energies $\hbar\omega$.
    }
\end{figure*}

To investigate the nonlinear response, we consider the intensity dependence of the excitation energy at the time when the light pulse finishes.
Figures~\ref{fig:bulk_intensity}(a--e) are logarithmic plots of the final excitation energy $E_{\rm ex}(t=\tau_p)$ as a function of $I_m$ at $T_e=0$, 0.2, 0.6, 1.0, and 1.4 eV, respectively, with $\hbar \omega=1.2$ eV.
The four lines correspond respectively to the cases of the pulse duration $\tau_p=10$, 20, 40, and 80 fs.
At $T_e=0$ [Fig.~\ref{fig:bulk_intensity}(a)], the four lines above $I_m=10^{11}$ W/cm$^2$ are proportional to $I_m^2$ and it indicates two-photon absorption as expected from the bandgap of about 2.4 eV and the photon energy of 1.2 eV.
Above $I_m=10^{11}$ W/cm$^2$, the excitation energy is roughly doubled when the pulse duration is doubled.
This implies that the excitation energy by the two-photon absorption is determined by the fluence of the light pulse.
The excitation energy below $I_m=10^{11}$ W/cm$^2$ is very small and decreases as the pulse duration increases.
This is because the below-gap absorption with lower intensities stems from high-frequency components of the applied pulse\cite{Yamada2019}.
The Fourier transform of the ultrashort pulse indicates a broadened spectrum whose tail decreases as the pulse duration increases.
The tail region around 2.4 eV causes weak one-photon excitation.
Also, the line of $\tau_p=80$ fs with lower intensities is proportional to $I_m^3$ (three-photon absorption).
This is because the calculated bandgap is slightly larger than 2.4 eV and the spectral tail of the $\tau_p=80$ fs pulse does not cause the excitation around 2.4 eV.

At $T_e=0.2$ eV [Fig.~\ref{fig:bulk_intensity}(b)], the behavior below $I_m=10^{11}$ W/cm$^2$ is drastically changed while the two-photon absorption region above that intensity is almost unchanged.
As expected from the dielectric function, the excitation energy below $10^{9}$ W/cm$^2$ exhibits the behavior of the one-photon linear absorption which is proportional to both the peak intensity $I_m$ and the pulse duration $\tau_p$. 
As the intensity increases, the lines gradually converge to a certain point.
Notably, the excitation energy at $I_m=5\times 10^{10}$ W/cm$^2$ has almost the same value regardless of the pulse duration.
At this point, the field energy cannot be absorbed over a threshold even with a longer pulse duration.
Namely, the saturable absorption occurs in the finite-temperature electron system even though it does not occur at zero temperature.
We shall call it the finite-temperature saturable absorption (FTSA).
Beyond FTSA, the lines gradually rise with the slope of the two-photon absorption.
We note that FTSA has not been recognized in existing numerical models for laser ablation of semiconductors, where absorption of laser pulses is assumed to be the sum of one-photon and two-photon processes\cite{Rethfeld2017}.
Following the assumption of such models, Fig.~\ref{fig:bulk_intensity}(b) should be a simple superposition of straight lines of the one-photon and two-photon processes.
This indicates that the existing models fail to predict the absorption of laser pulses under certain conditions.

In Figs.~\ref{fig:bulk_intensity}(c--e), we can see similar behavior but the intensity of FTSA slightly increases as the electron temperature increases. 
For $T_e=1$ eV [Fig.~\ref{fig:bulk_intensity}(d)], FTSA occurs at $I_m=2\times 10^{11}$ W/cm$^2$.
The characteristics of FTSA become more clear at high temperatures $T_e\ge 0.6$ eV.
As the intensity $I_m$ increases, the one-photon linear absorption below $10^{9}$ W/cm$^2$ transitions into the FTSA region, in which the excitation energy is saturated. 
The saturated excitation energy is lower than that of the expected linear absorption.
As increasing $I_m$ further, the two-photon absorption becomes dominant over FTSA.

Figure~\ref{fig:bulk_intensity}(f) presents results with $T_e=1$ eV and $\tau_p=80$ fs for several photon energies of $\hbar \omega=0.8$, 1.2, and 1.6 eV.
Even changing the photon energy, the overall behavior of the excitation energy is almost unchanged.
Hereafter, we fix the photon energy as $\hbar \omega=1.2$ eV throughout this paper.
The above observations suggest that FTSA is a universal phenomenon to some extent.

\begin{figure}
    \includegraphics[keepaspectratio,width=\columnwidth]
    {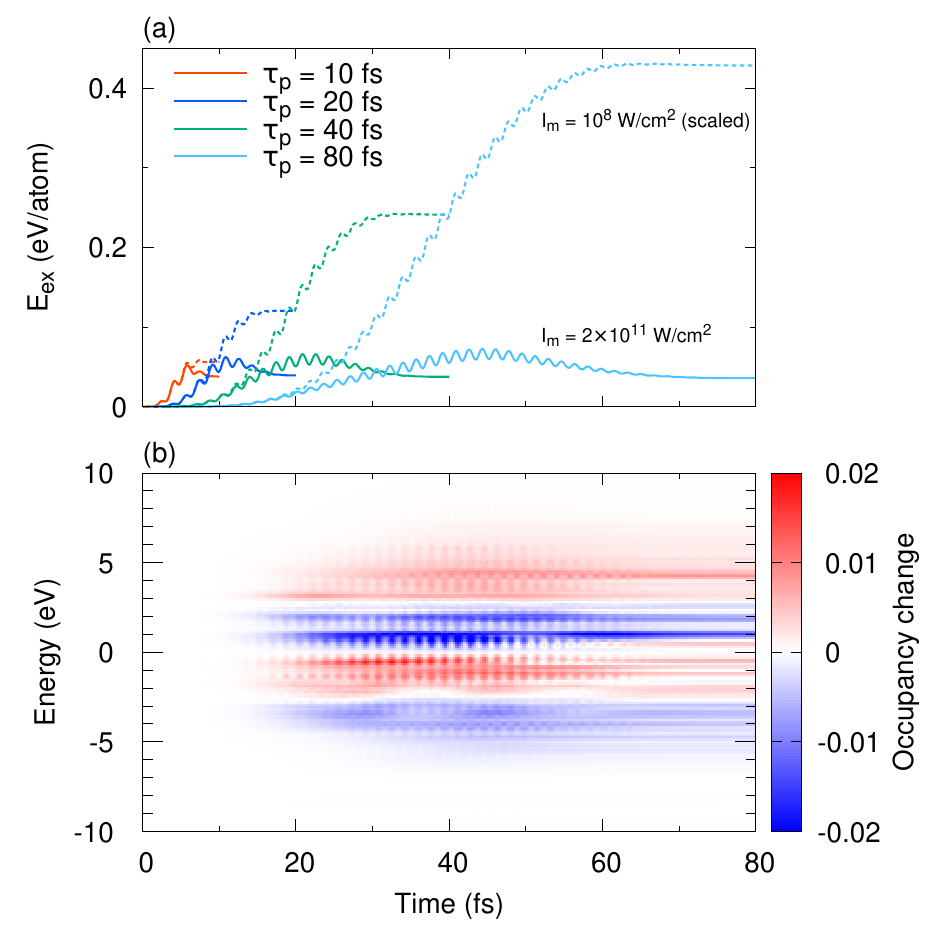}
    \caption{\label{fig:bulk_time}
    (a) Time profile of the excitation energy at $T_e=1$ eV with several values of the pulse duration $\tau_p$.
    The solid lines are for the case of $I_m=2\times 10^{11}$ W/cm$^2$. 
    The dashed lines are for $I_m=10^{8}$ W/cm$^2$ but scaled up by a factor of $2\times10^3$ for comparison.
    (b) Occupancy change $\tilde{D}(\varepsilon,t)/D(\varepsilon)$ [Eq.~(\ref{eq:pdos})] as a function of the time $t$ and the orbital energy $\varepsilon$ for the case of $\tau_p=80$ fs, $T_e=1$ eV, and $I_m=2\times 10^{11}$ W/cm$^2$.
    The chemical potential is at $\varepsilon=0$.
    }
\end{figure}

To obtain physical insights into FTSA, we shall pick up the case of $T_e=1$ eV and $I_m=2\times 10^{11}$ W/cm$^2$.
Solid lines in Fig.~\ref{fig:bulk_time}(a) are the excitation energy as a function of the time for that case with the respective $\tau_p$. 
Dashed lines are the excitation energy with $I_m=10^{8}$ W/cm$^2$ (linear regime) but scaled up by a factor of $2\times10^3$ for comparison.
The dashed lines exhibit the usual behavior of the one-photon linear absorption.
At the beginning of the pulse, the solid and dashed lines are equivalent but the difference of the two becomes apparent as the time increases by nonlinear interaction.
The solid lines finally reach a value of 0.036--0.039 eV/atom, which is the saturated excitation energy.
The excitation energy at FTSA is significantly reduced from the value expected by the linear absorption.

Fig.~\ref{fig:bulk_time}(b) shows the occupancy change $\tilde{D}(\varepsilon,t)/D(\varepsilon)$ [Eq.~(\ref{eq:pdos})] as a function of the time $t$ and the orbital energy $\varepsilon$ for the case of $\tau_p=80$ fs.
The upper (lower) half corresponds to the conduction (valence) band, where the energy level $\varepsilon=0$ is set to the chemical potential. 
The absolute value of the occupancy change reaches the maximum around the middle point of the pulse, $t=40$ fs, and becomes constant after $t\sim 65$ fs.
The occupancy change is positive above $\varepsilon\sim 3$ eV ($\sim -3$ eV) for the conduction (valence) band but negative below.
Because the two-photon absorption across the bandgap is negligible with $I_m=2\times 10^{11}$ W/cm$^2$, the conduction and valence bands are separable. 
This implies that the one-photon transitions of conduction-to-conduction and valence-to-valence are saturated respectively.

\begin{figure}
    \includegraphics[keepaspectratio,width=\columnwidth]
    {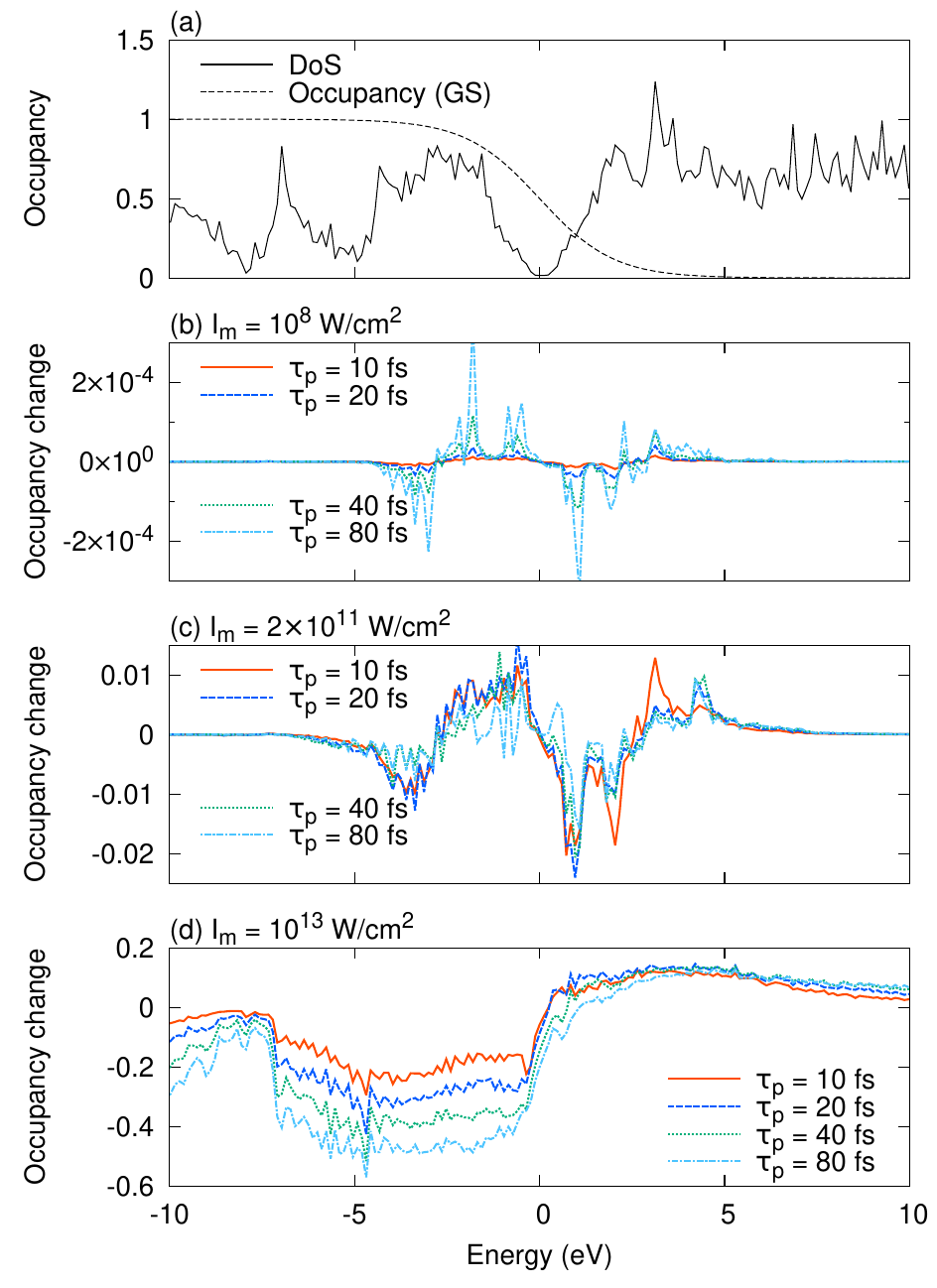}
    \caption{\label{fig:bulk_occ}
    (a) Density of states $D(\varepsilon)$ and electron occupancy at $T_e=1$ eV.
    (b--d) Occupancy changes at the pulse end with the intensities of $I_m=10^8$, $2\times 10^{11}$, and $10^{13}$ W/cm$^2$, respectively, with several values of the pulse duration $\tau_p$.
    }
\end{figure}

In Fig.~\ref{fig:bulk_occ}(a), the density of states $D(\varepsilon)$ and the occupancy in the ground state at $T_e=1$ eV are illustrated. 
Figures~\ref{fig:bulk_occ}(b--d) present the occupancy changes at the pulse end $t=\tau_p$ with the intensity $I_m=10^8$, $2\times 10^{11}$, and $10^{13}$ W/cm$^2$, respectively.
In the linear regime [Fig.~\ref{fig:bulk_occ}(b)], the absolute value of the occupancy change increases monotonically as the pulse duration $\tau_p$ increases due to one-photon transitions in the conduction or valence band.
At the FTSA intensity [Fig.~\ref{fig:bulk_occ}(c)], the occupancy change seems to be almost saturated, where the light-blue dash-dotted line corresponds to the vertical line at $t=80$ fs in Fig.~\ref{fig:bulk_time}(b).
For the very high intensity [Fig.~\ref{fig:bulk_occ}(d)], the occupancy in the valence band decreases monotonically as $\tau_p$ increases because of the two-photon absorption across the bandgap.
The change in the conduction band seems to be small but this may be because the change in the higher energy region is dominant.
In any case, FTSA occurs when the occupation changes by the one-photon transitions of conduction-to-conduction and valence-to-valence are saturated and the two-photon absorption across the bandgap is negligible.

\subsection{Toy model analysis\label{sec:toy}}

\begin{figure}
    \begin{tabular}{c}
    \includegraphics[keepaspectratio,width=\columnwidth]{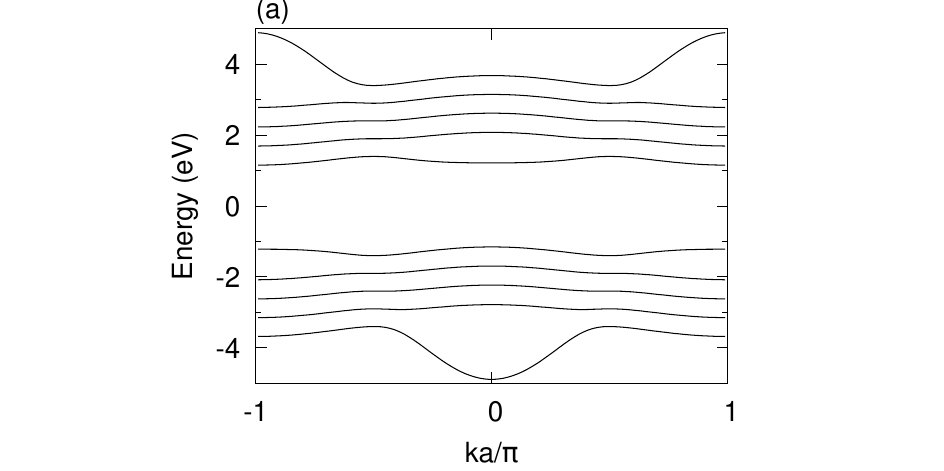}\\
    \includegraphics[keepaspectratio,width=\columnwidth]{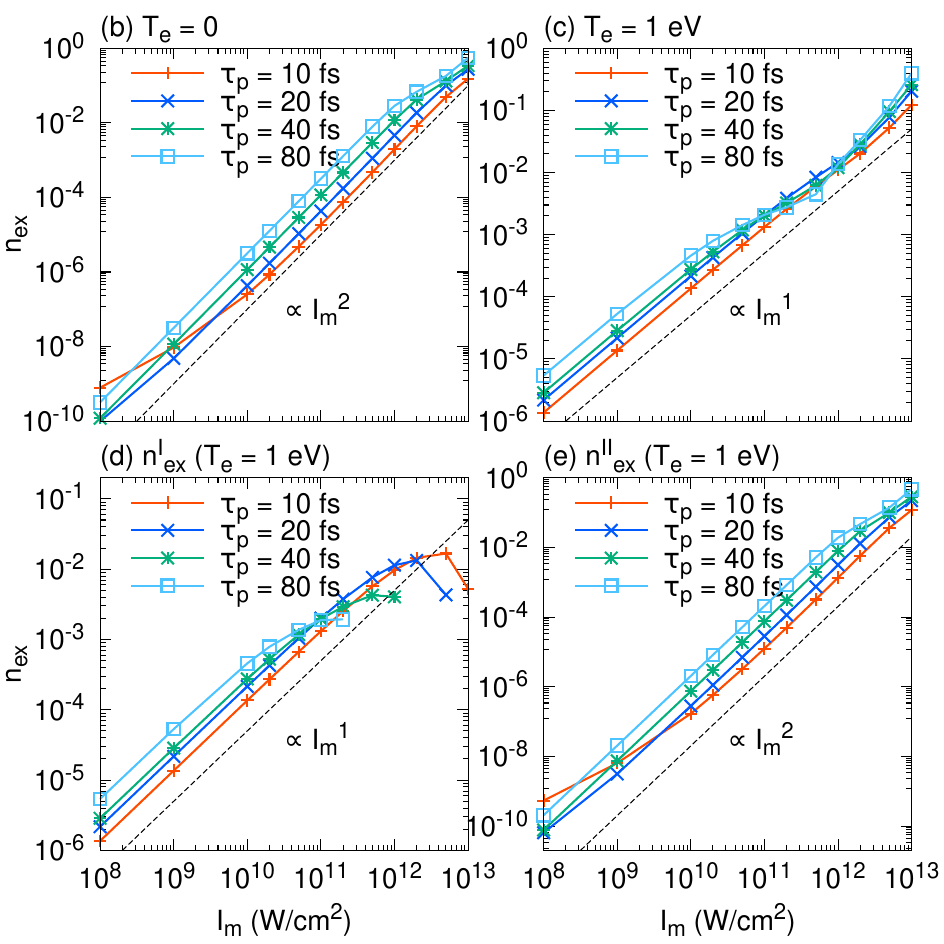}
    \end{tabular}
    \caption{\label{fig:model}
    (a) Band structure of the toy model.
    (b,c) The same as Fig.~\ref{fig:bulk_intensity}(a,d), respectively, but for the toy model.
    The vertical axis is for the excited carrier density.
    (d,e) One-photon and two-photon parts, respectively, of the panel (c).
    }
\end{figure}

From the above discussions, FTSA may be general in gapped multiple-band systems at finite temperatures and originate from the saturation of the electron occupancy.
However, it is difficult to identify the cause of the occupancy saturation from the above results because the TDDFT-based first-principles calculations incorporate enormous physical effects.
To confirm the physical mechanisms behind FTSA, we shall simplify the problem by shaving unnecessary factors.
In this subsection, we consider the toy model in Sec.~\ref{sec:method_toy} instead of Si.
Here we employ a ten-band model, $N_{\rm band}=10$, where the conduction (valence) band comprises upper (lower) 5 bands.
The parameters of the model are set as $\varepsilon_{6}-\varepsilon_5 =$ 2.8 eV, $\varepsilon_{b+1}-\varepsilon_b = $ 0.5 eV, ($b=1,\cdots,4,6,\cdots,9$), $\tilde{t}=0.2$ eV, $a = 2.5$ {\AA}, and $N=100$.
The band structure at the ground state is depicted in Fig.~\ref{fig:model}(a).
This toy model has a bandgap of around 2.4 eV and highly abstracted properties of bulk Si.

Figures~\ref{fig:model}(b,c) are analogs of Figs.~\ref{fig:bulk_intensity}(a,d), respectively.
That is, Figs.~\ref{fig:model}(b,c) show the excited carrier densities $n_{\rm ex}(t=\tau_p)$ [Eq.~(\ref{eq:nex_model})] as a function of the intensity at the temperature $T_e=0$ and 1 eV, respectively.
The overall behaviors are qualitatively the same as before and FTSA occurs in Fig.~\ref{fig:model}(c).
We confirmed that the case of $N_{\rm band}=2$ does not exhibit FTSA.
Note that FTSA never occurs in a gapped two-band model because the one-photon absorption in the respective conduction and valence bands cannot be considered in such a system.
FTSA is essentially due to a multiple-band effect.

Furthermore, we separate the contributions from the one-photon and two-photon absorption in the excited carrier density, $n_{\rm ex}(t)=n_{\rm ex}^{\rm I}(t)+n_{\rm ex}^{\rm II}(t)$, as follows: 
\begin{eqnarray}
    n_{\rm ex}^{\rm I}(t) &\equiv& 
    \frac{1}{N}\sum_k\sum_{c,c'} 
    f_{ck} (1-f_{c'k})
    |\langle \psi^{\rm GS}_{c'k} | 
    \psi_{ck}(t)\rangle |^2 \nonumber \\
    &+&
    \frac{1}{N}\sum_k\sum_{v,v'} 
    f_{vk} (1-f_{v'k})
    |\langle \psi^{\rm GS}_{v'k} | 
    \psi_{vk}(t)\rangle |^2 ,
    \\ 
    n_{\rm ex}^{\rm II}(t) &\equiv&
    \frac{1}{N}\sum_k\sum_{c,v} 
    f_{ck} (1-f_{vk})
    |\langle \psi^{\rm GS}_{vk} | 
    \psi_{ck}(t)\rangle |^2 \nonumber \\
    &+&
    \frac{1}{N}\sum_k\sum_{c,v} 
    f_{vk} (1-f_{ck})
    |\langle \psi^{\rm GS}_{ck} | 
    \psi_{vk}(t)\rangle |^2,
\end{eqnarray}
where $c$ and $v$ are band indices for the conduction and valence bands, respectively.
$n_{\rm ex}^{\rm I}(t)$ represents the one-photon contribution of conduction-to-conduction and valence-to-valence transitions.
$n_{\rm ex}^{\rm II}(t)$ corresponds to the two-photon contribution across the bandgap.
Figures~\ref{fig:model}(d,e) show $n_{\rm ex}^{\rm I}(t=\tau_p)$ and $n_{\rm ex}^{\rm II}(t=\tau_p)$, respectively, at $T_e=1$ eV.
The two-photon contribution [Fig.~\ref{fig:model}(e)] is very similar to the zero-temperature result [Fig.~\ref{fig:model}(b)] as expected.
For the one-photon contribution [Fig.~\ref{fig:model}(d)], the lines reach a ceiling at high intensities.
This is because of the Pauli blocking by the strong one-photon absorption, where further one-photon transition is prohibited by the Pauli exclusion principle by existing excited carriers.
While such one-photon transition cannot contribute to $n_{\rm ex}(t)$ at zero temperature ($f_{vk}=1$ and $f_{ck}=0$), it is dominant for lower intensities at finite temperatures.
From the above discussion, we confirmed that FTSA originates from the saturation of the one-photon absorption corresponding to the conduction-to-conduction and valence-to-valence transitions.

\section{Surface excitation and light propagation\label{sec:results2}}

\subsection{TDDFT results for Si slab \label{sec:tddft_slab}}

We consider the same calculations as in Sec.~\ref{sec:tddft_bulk} but for Si slab systems.
Figures~\ref{fig:lr}(c,d) show the $xx$ component of the dielectric function for the Si slab with the thickness of $d=10a=5.43$ nm. 
Here, to compare with Figs.~\ref{fig:lr}(a,b), the slab dielectric function is converted to a value for the 3D material by considering the slab thickness.
That is converted from the conductivity for the 3D material, which is equal to the 2D conductivity of the slab divided by the thickness\cite{Yamada2018}.
The dielectric functions of the bulk [Figs.~\ref{fig:lr}(a,b)] and slab [Figs.~\ref{fig:lr}(c,d)] are almost equivalent to each other except that the peak of the imaginary part at around 4 eV is slightly reduced in the slab case.
The dielectric function is almost unchanged by the slab surfaces.

\begin{figure}
    \begin{tabular}{c}
    \includegraphics[keepaspectratio,width=\columnwidth]{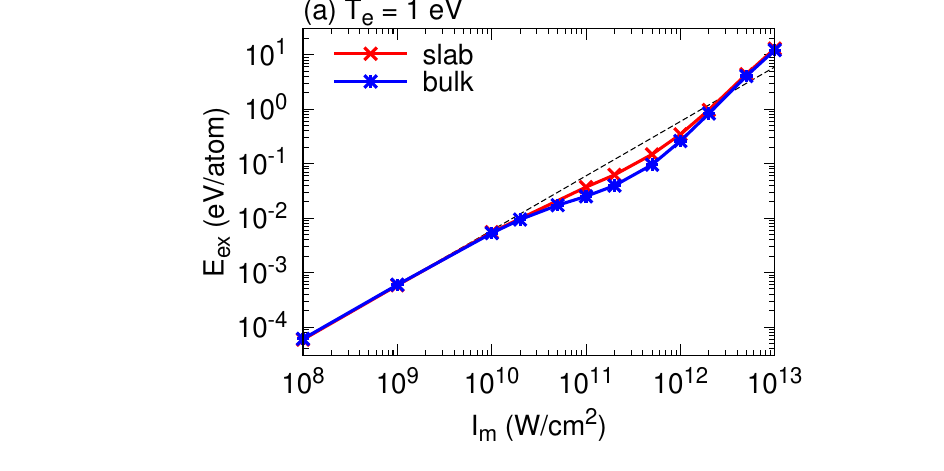}\\
    \includegraphics[keepaspectratio,width=\columnwidth]{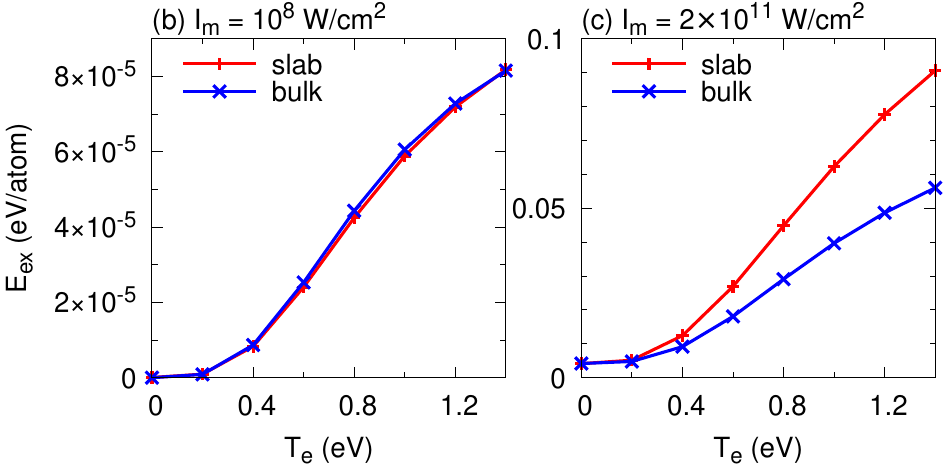}
    \end{tabular}
    \caption{\label{fig:slab}
    Comparison of the excitation energy per Si atom between the bulk and $d=10a$ slab results.
    (a) Intensity dependence at $T_e=1$ eV.
    (b,c) Temperature dependence at $I_m=10^8$ and $2\times 10^{11}$ W/cm$^2$, respectively.
    }
\end{figure}

We shall consider the intensity dependence of the final excitation energy $E_{\rm ex}(t=\tau_p)$ per Si atom in the $d=10a$ slab.
Figure~\ref{fig:slab}(a) shows the comparison between the results of the bulk and slab ($d=10a$) cases at $T_e=1$ eV and $\tau_p= 20$ fs.
As mentioned above, the photon energy is fixed to $\hbar \omega=1.2$ eV throughout this section.
The black dashed line represents the slope of the linear absorption.
While the solid lines of the bulk and slab cases coincide with each other below $I_m=10^{10}$ W/cm$^2$ and above $5\times 10^{12}$ W/cm$^2$, the slab value is greater than the bulk value around the intensity FTSA occurs.
Note that both the bulk and slab values around FTSA intensity are lower than that expected from the linear excitation (black dashed line).

Figures~\ref{fig:slab}(b) and (c) present the final excitation energies as a function of $T_e$ for the cases of $I_m=10^8$ and $2\times 10^{11}$ W/cm$^2$, respectively, with the same pulse duration and photon energy.
In the former case, the linear excitation is not affected by the surfaces at all the temperatures as expected from the dielectric function in Fig.~\ref{fig:lr}.
In the latter case (FTSA), on the other hand, the slab energy is amplified from the bulk energy as the temperature increases.
At $T_e=0$, such amplification by the surfaces does not occur.
The above results at $T_e=0$ are consistent with our previous works which have shown that the surface effect on the electronic excitation in a relatively thick slab ($d > 5$ nm) is negligible irrespective of the field strength at zero temperature\cite{Yamada2018,Yamada2021}.

\begin{figure}
    \includegraphics[keepaspectratio,width=\columnwidth]
    {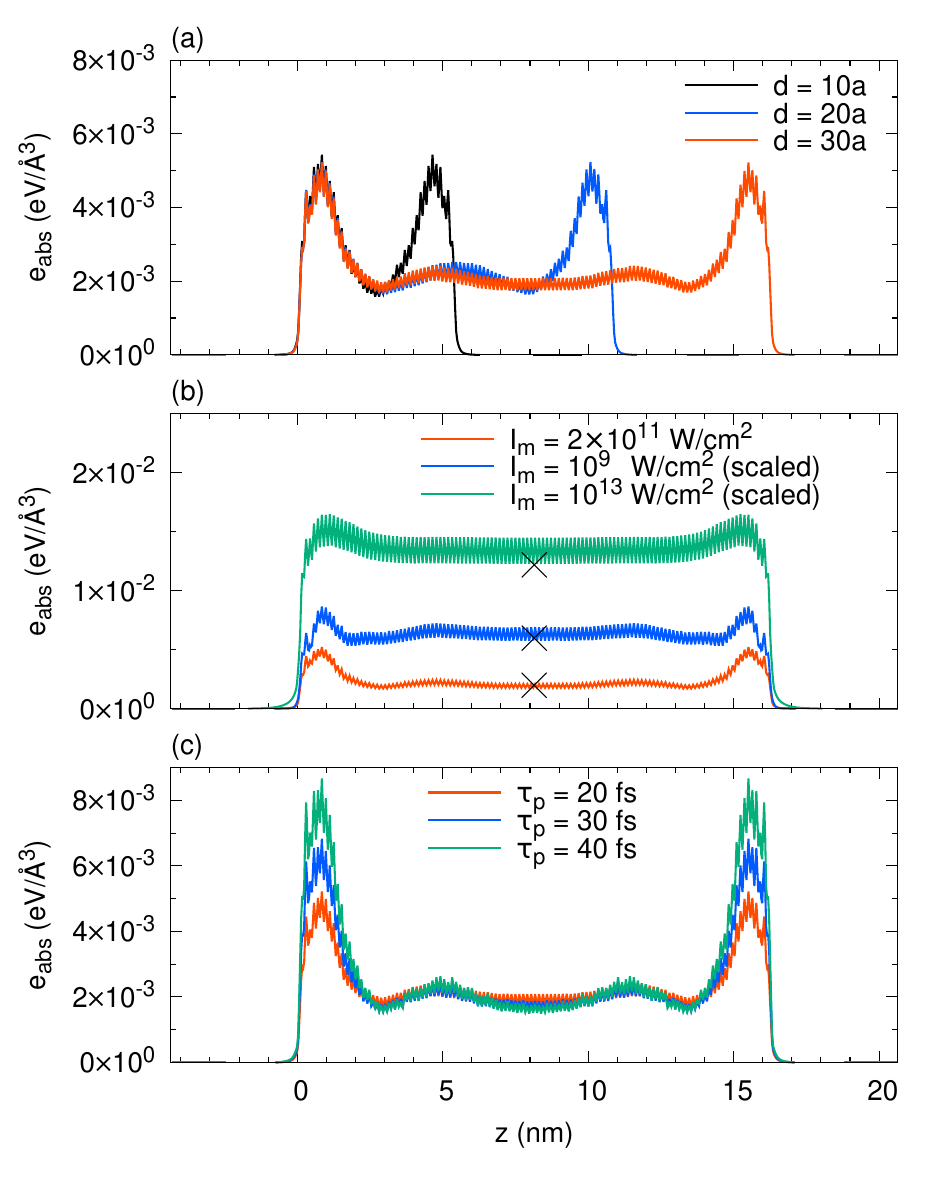}
    \caption{\label{fig:slab_eabs}
    Absorbed energy density [Eq.~(\ref{eq:eabs})] of Si slabs at $T_e=1$ eV as a function of $z$.
    (a) Results for the thicknesses of $d=10a$, $20a$, and $30a$, with $I_m=2\times 10^{11}$ W/cm$^2$ and $\tau_p= 20$ fs.
    (b) Results for several intensities of $I_m=2\times 10^{11}$, $10^9$, and $10^{13}$ W/cm$^2$ with $d=30a$ and $\tau_p= 20$ fs. 
    The results of  $I_m=$ $10^9$ and $10^{13}$ W/cm$^2$ are multiplied by a scaling factor assuming linear absorption  (see the text for details).
    The ``$\times$" symbols represent the corresponding results by the bulk calculations.
    (c) Results with $\tau_p= 20$, 30, and 40 fs, where $d=30a$ and $I_m=2\times 10^{11}$ W/cm$^2$.
    }
\end{figure}

To look into the absorption enhancement by the surfaces, we observe the absorbed energy density [Eq.~(\ref{eq:eabs})] at $T_e=1$ eV.
Figure~\ref{fig:slab_eabs} shows results for the final absorbed energy density, $e_{\rm abs}({\bf r},t=\tau_p)$, averaged over the $xy$ plane as a function of $z$, where the front surface of the slab is located at $z=0$.
Figure~\ref{fig:slab_eabs}(a) provides results for the thicknesses of $d=10a$, $20a$, and $30a$, with the FTSA intensity $I_m=2\times 10^{11}$ W/cm$^2$ and the pulse duration $\tau_p= 20$ fs.
In Fig.~\ref{fig:slab_eabs}(b), we compare results for several intensities of $I_m=2\times 10^{11}$, $10^9$, and $10^{13}$ W/cm$^2$ with $d=30a$ and $\tau_p= 20$ fs, where the latter two results are multiplied by a scaling factor assuming linear absorption to correspond to the former.
The scaling factor for the case of $I_m=10^9$ W/cm$^2$ ($10^{13}$ W/cm$^2$) is equal to $2\times 10^{11}/10^9=200$ ($2\times 10^{11}/10^{13}=0.02$).
If the absorption is linear, the three lines should be equivalent to each other.
The ``$\times$" symbols in Fig.~\ref{fig:slab_eabs}(b) represent the corresponding results by the bulk calculations, which are comparable to the respective solid lines at the middle point of the slab.
Figure~\ref{fig:slab_eabs}(c) shows results with 
$\tau_p= 20$, 30, and 40 fs, where $d=30a$ and $I_m=2\times 10^{11}$ W/cm$^2$.
Throughout Fig.~\ref{fig:slab_eabs}, the case of $d=30a$, $I_m=2\times 10^{11}$ W/cm$^2$, and $\tau_p= 20$ fs is plotted as red solid line.
In the entire region, the absorbed energy density shows fine oscillations with a period of about 0.1 nm, which stem from the Si-Si bond.
There are four periods per the lattice constant $a=0.543$ nm.

In Fig.~\ref{fig:slab_eabs}(a), the absorption is enhanced in regions from the surface to a depth of around 2.5 nm.
This enhancement region is unchanged by increasing the thickness $d$.
In the bulk region deeper than 2.5 nm from the surfaces, the absorption is roughly flat and coincides with the bulk result.

In Fig.~\ref{fig:slab_eabs}(b), the enhancement near the surfaces is not apparent for the low and very high intensities.
For the low intensity (blue solid line), the modification of the linear absorption near the surfaces originates from the surface electronic structure\cite{Yamada2018}.
For the very high intensity (green solid line), the absorption at the entire region is enhanced from that expected from the linear response because the two-photon absorption is dominant.
At the FTSA intensity (red solid line), the absorption in the bulk region is greatly reduced from the linear absorption due to FTSA.
Near the surfaces, the FTSA reduction is moderate compared to the bulk region and it causes relative enhancement of the absorption.

By the $\tau_p$ dependence in Fig.~\ref{fig:slab_eabs}(c), we can see that the absorption near the surfaces is proportional to the fluence while that in the bulk region is saturated.
Using longer pulses with the FTSA intensity, the surface region continuously absorbs the energy from the electromagnetic field but the bulk region does not.

From Fig.~\ref{fig:slab} and Fig.~\ref{fig:slab_eabs}, we conclude that the surface of Si relatively enhances the absorption near the surface compared to that in the bulk region.
This surface-induced relative enhancement is actually just the suppression of FTSA and the enhanced absorption does not exceed the scaled linear absorption.
The enhancement occurs only when the following conditions are satisfied: (1) finite temperature, (2) moderately high-intensity light, (3) surface.

\subsection{Dynamics of excitation energy density}

\begin{figure}
    \includegraphics[keepaspectratio,width=\columnwidth]{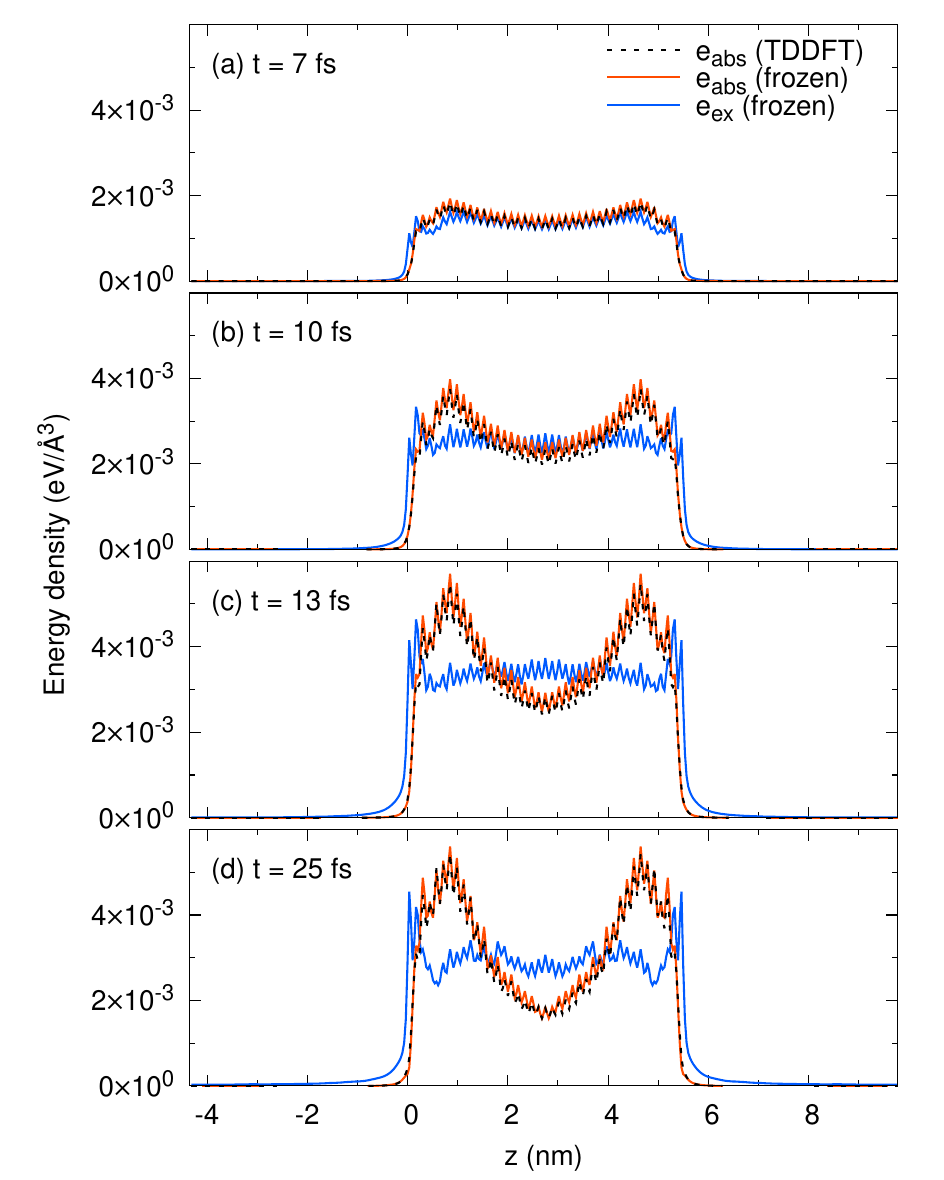}
    \caption{\label{fig:slab_frozen}
    Snapshots of the energy density at each time with $T_e=1$ eV, $d=10a$, $\tau_p=20$ fs, and $I_m=2\times 10^{11}$ W/cm$^2$.
    The black dashed line (red solid line) is the absorbed energy density by TDDFT (frozen Hamiltonian formalism).
    The blue solid line indicates the excitation energy density by the frozen Hamiltonian formalism.
    }
\end{figure}

From Fig.~\ref{fig:slab_eabs}(c), one may deduce that electrons near the surface hold higher energy than that in the bulk region even after the pulse end, but this is not the case.
From the definition, the absorbed energy density [Eq.~(\ref{eq:eabs})] just represents a history of the distribution of the energy transfer from the electromagnetic field to the electron system.
The migration of the electronic energy density inside the material after absorption is not recorded in the absorbed energy density.

To trace the migration of the electronic energy, we define the excitation energy density of electrons as
\begin{eqnarray}
    e_{\rm ex}({\bf r},t)&=& \frac{1}{N_k}\sum_{{\bf k},n}f_{n{\bf k}} \,{\rm Re}\left[ u^{\ast}_{n{\bf k}}({\bf r},t) \,\hat{H}^{\rm KS}_{\bf k}(t) 
\, u_{n{\bf k}}({\bf r},t) \right] \nonumber \\
&& - ({\rm GS \,\,value}),
\end{eqnarray}
where the latter term is the value of the former term at $t=0$.
In TDDFT, the spatial integral of $e_{\rm ex}({\bf r},t)$ disagrees with the total excitation energy because the Hartree and exchange-correlation terms are inhomogeneous with respect to the electron density $n_{\rm e}({\bf r},t)$.
For this reason, we shall use the frozen Hamiltonian formalism in which the Hartree and exchange-correlation potentials in the one-electron Hamiltonian $\hat{H}^{\rm KS}_{\bf k}(t)$ are fixed.
In this formalism, Eq.~(\ref{eq:tdks}) is equivalent to the time-dependent Schroedinger equation driven by $\hat{H}^{\rm KS}_{\bf k}(t)$ with the fixed ground-state Kohn-Sham potential.
In the frozen Hamiltonian formalism, the spatial integral of $e_{\rm ex}({\bf r},t)$ is equal to the total excitation energy $E_{\rm ex}(t)\equiv E_{\rm tot}(t)-E_{\rm tot}(t=0)$ [see Eq.~(\ref{eq:total_energy})].
According to the energy conservation, the spatial integrals of $e_{\rm ex}({\bf r},t)$ and $e_{\rm abs}({\bf r},t)$ are equivalent to each other.

Figure~\ref{fig:slab_frozen} provides snapshots of the energy density at each time with $T_e=1$ eV, $d=10a$, $\tau_p=20$ fs, and $I_m=2\times 10^{11}$ W/cm$^2$ [the same conditions as the black solid line in Fig.~\ref{fig:slab_eabs}(a)].
The black dashed line and red solid line represent the $xy$-averaged absorbed energy densities, $e_{\rm abs}(z,t)$, for TDDFT and frozen Hamiltonian formalism, respectively.
They are almost equivalent to each other and it indicates that the time-dependence of the local potential in TDDFT has little effect.
The blue solid line corresponds to the excitation energy density averaged over the $xy$ plane, $e_{\rm ex}(z,t)$.
At $t=7$ fs [Fig.~\ref{fig:slab_frozen}(a)], $e_{\rm ex}(z,t)$ almost coincides with $e_{\rm abs}(z,t)$ but the surface region has small deviation.
After that, the  relative enhancement of $e_{\rm abs}(z,t)$ at the surfaces gradually grows but $e_{\rm ex}(z,t)$ is roughly flat in the slab and slightly spills out into the vacuum.
While $e_{\rm ex}(z,t)$ has sharp spikes near the surfaces due to the surface electronic structure, its entire shape is roughly rectangular.
This indicates that the energy density transferred from the electromagnetic field to the electron system immediately spreads out the entire slab and the electronic excitation energy density is roughly averaged in the slab.
If we irradiate the surface of a semi-infinite solid with a very long light pulse under FTSA conditions, it is expected that the energy absorbed at the surface region  continuously flows into the bulk region during light irradiation.

\subsection{Light propagation\label{sec:light_propagation}}

\begin{figure}
    \includegraphics[keepaspectratio,width=\columnwidth]
    {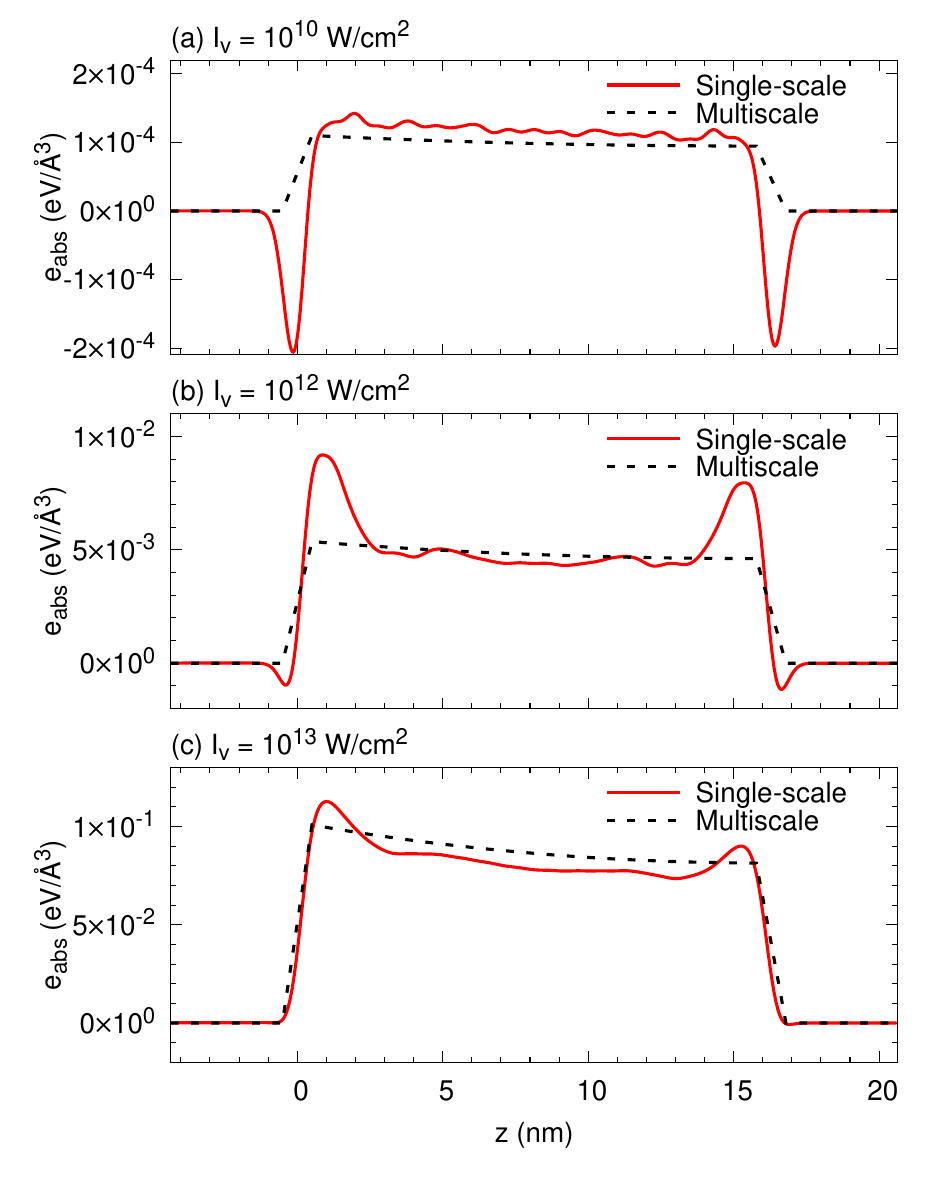}
    \caption{\label{fig:ms_ss}
    Comparison of the absorbed energy density between the single-scale and multiscale Maxwell-TDDFT methods with $T_e=1$ eV, $d=30a$, and $\tau_p=20$ fs.
    (a--c) Results for low-intensity ($I_v=10^{10}$ W/cm$^2$), high-intensity ($I_v=10^{12}$ W/cm$^2$), and very high-intensity ($I_v=10^{13}$ W/cm$^2$) cases, respectively.
    }
\end{figure}

To consider the correspondence between TDDFT calculations and experiments, we evaluate not only the surface effect but also the light propagation effect.
We shall assess the applicability of the two Maxwell-TDDFT methods and the validity of the coarse-graining approximation (Sec.~\ref{sec:method_light_propagation}).
In the Maxwell-TDDFT methods, the incident pulse in the vacuum at $t=0$ is prepared at the front of the Si surface ($z=0$).
Thus, the field strength of the pulse is specified by the peak intensity in the vacuum, $I_v$, rather than that in the medium, $I_m$.

Figure~\ref{fig:ms_ss} provides a comparison of the $xy$-averaged absorbed energy density between the two Maxwell-TDDFT methods with $T_e=1$ eV, $d=30a$, and $\tau_p=20$ fs.
Figures~\ref{fig:ms_ss}(a--c) represent results for low-intensity ($I_v=10^{10}$ W/cm$^2$), high-intensity ($I_v=10^{12}$ W/cm$^2$), and very high-intensity ($I_v=10^{13}$ W/cm$^2$) cases, respectively.
We note that results by the two methods for Si films with $d>5$ nm at zero-temperature are almost equivalent to each other irrespective of the field strength of incident pulses\cite{Yamada2018,Yamada2021}.
The absorbed energy density by the single-scale method is calculated by the surface integral of the Poynting vector over the $xy$ plane at each $z$, which is equivalent to Eq.~(\ref{eq:eabs}) with the microscopic electric field.
We have applied the Gaussian filter with a width of 10 a.u.\ on the single-scale results to suppress their spiky noises due to the  microscopic treatment of the electromagnetic field. 

For the low-intensity case in Fig.~\ref{fig:ms_ss}(a), the single-scale result has two dips at the surfaces.
The dips are apparent in lower intensity cases only and originate from the microscopic fluctuations of the current density and electric field at the surfaces\cite{Yamada2018}.
Such microscopic fluctuations are ignored in the slab TDDFT calculations, in which the electric field is spatially uniform. 
While the contribution of the microscopic fluctuations to the optical response is important when $d<5$ nm, it is negligible for thicker slabs of $d>5$ nm\cite{Yamada2018}.
Except for the dips, the single-scale and multiscale results agree with each other.
Due to light propagation in the medium, the energy absorption at the front edge is slightly greater than that at the back edge.

In Fig.~\ref{fig:ms_ss}(b), the surface-induced enhancement with FTSA is apparent in the regions from the surface to a depth of around 2.5 nm for the single-scale case.
The multiscale method fails to reproduce the surface-induced enhancement because it cannot account for the surface electronic structure.
Nevertheless, the single-scale result in the bulk region is well reproduced by the multiscale method.
This suggests that the multiscale method can properly calculate light propagation even with FTSA except for the surface-induced enhancement.

For the very high-intensity case in Fig.~\ref{fig:ms_ss}(c), the single-scale and multiscale results agree well with each other again.
The overall behaviors in Fig.~\ref{fig:ms_ss} are comparable to those in Fig.~\ref{fig:slab_eabs}(b) for the low-intensity, high-intensity (FTSA), and very high-intensity cases, respectively.
Even considering light propagation, the surface-induced enhancement with FTSA occurs in single-scale Maxwell-TDDFT calculations as in the slab TDDFT calculations.

The single-scale Maxwell-TDDFT method enables us to directly simulate laser-irradiation experiments for FTSA including the surface and light propagation effects but requires enormous computational costs.
The multiscale Maxwell-TDDFT method is a powerful tool for simulating the macroscopic light propagation with FTSA because it can properly reproduce single-scale results except for microscopic regions near surfaces and its computational cost is moderate.
The effect of FTSA on the macroscopic light propagation in a finite-thickness slab, such as the penetration depth or modulations of the reflectance and transmittance, is a very interesting topic but beyond our scope here.
For zero-temperature systems, there are numerical studies that discussed the penetration depth and reflectance/transmittance modulations by the ordinary saturable absorption using the multiscale Maxwell-TDDFT method\cite{Uemoto2021,Yamada2024}.
Such calculations for FTSA will be useful for predicting future experimental measurements for this topic.

We just mention a calculation procedure for a semi-infinite solid with FTSA.
A rough estimate of the absorption near the surface can be established by the ordinary TDDFT method and it  is not necessarily required to use the multiscale Maxwell-TDDFT method.
Because $I_v$ corresponds to a parameter of the laser intensity used in experiments, $I_m$ should be converted from that.
At the surface of a semi-infinite solid, assuming the linear propagation, the light intensity in the medium $I_m$ is given by the vacuum intensity $I_v$ as 
\begin{equation}
    I_m = \left\vert \frac{2}{1+\sqrt{\varepsilon(\omega)}}\right\vert^2 I_v,
\end{equation}
where $\varepsilon(\omega)$ is the dielectric function for the bulk region as calculated in Figs.~\ref{fig:lr}(a,b).
From this relation, we can roughly estimate $I_m$ for the desired $I_v$ at each temperature $T_e$ even with high intensities.
Using the estimated $I_m$, the absorbed energy in FTSA can be evaluated by a unit-cell calculation for the bulk region near the surface as in Sec.~\ref{sec:tddft_bulk}.
With a slab calculation as in Sec.~\ref{sec:tddft_slab}, the surface effect on FTSA can also be calculated with $I_m$.

\subsection{Bulk Si under the surface-mimicking external potential}

\begin{figure}
    \begin{tabular}{c}
    \includegraphics[keepaspectratio,width=\columnwidth]{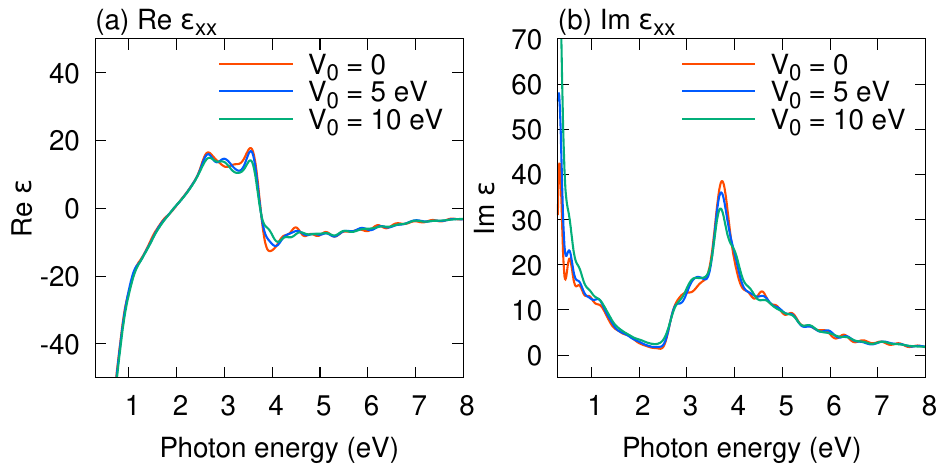}\\
    \includegraphics[keepaspectratio,width=\columnwidth]{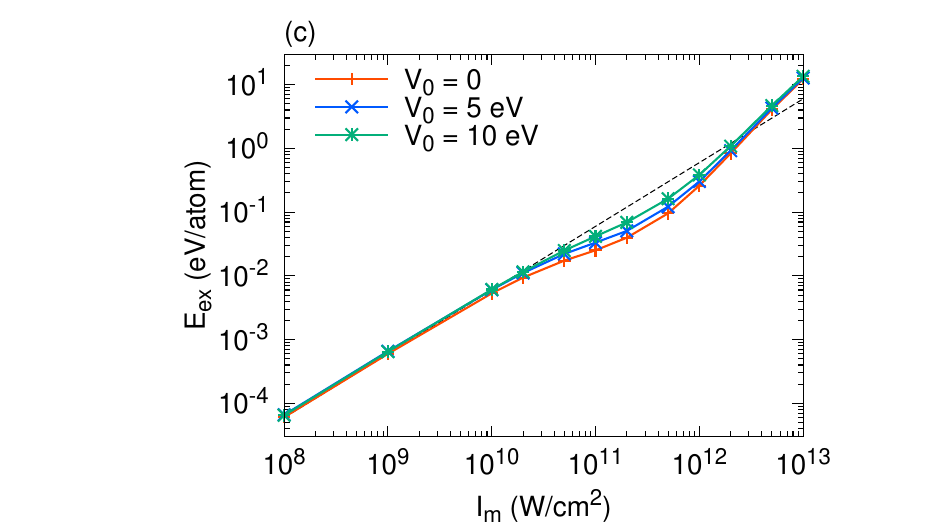}
    \end{tabular}
    \caption{\label{fig:ext_lr}
    (a) Real and (b) imaginary parts of the dielectric function at $T_e=1$ eV under the static external potential with several strengths of $V_0=0$, 5, and 10 eV.
    (c) Intensity dependence of the excitation energy at $T_e=1$ eV under the static external potential.
    }
\end{figure}

\begin{figure}
    \includegraphics[keepaspectratio,width=\columnwidth]
    {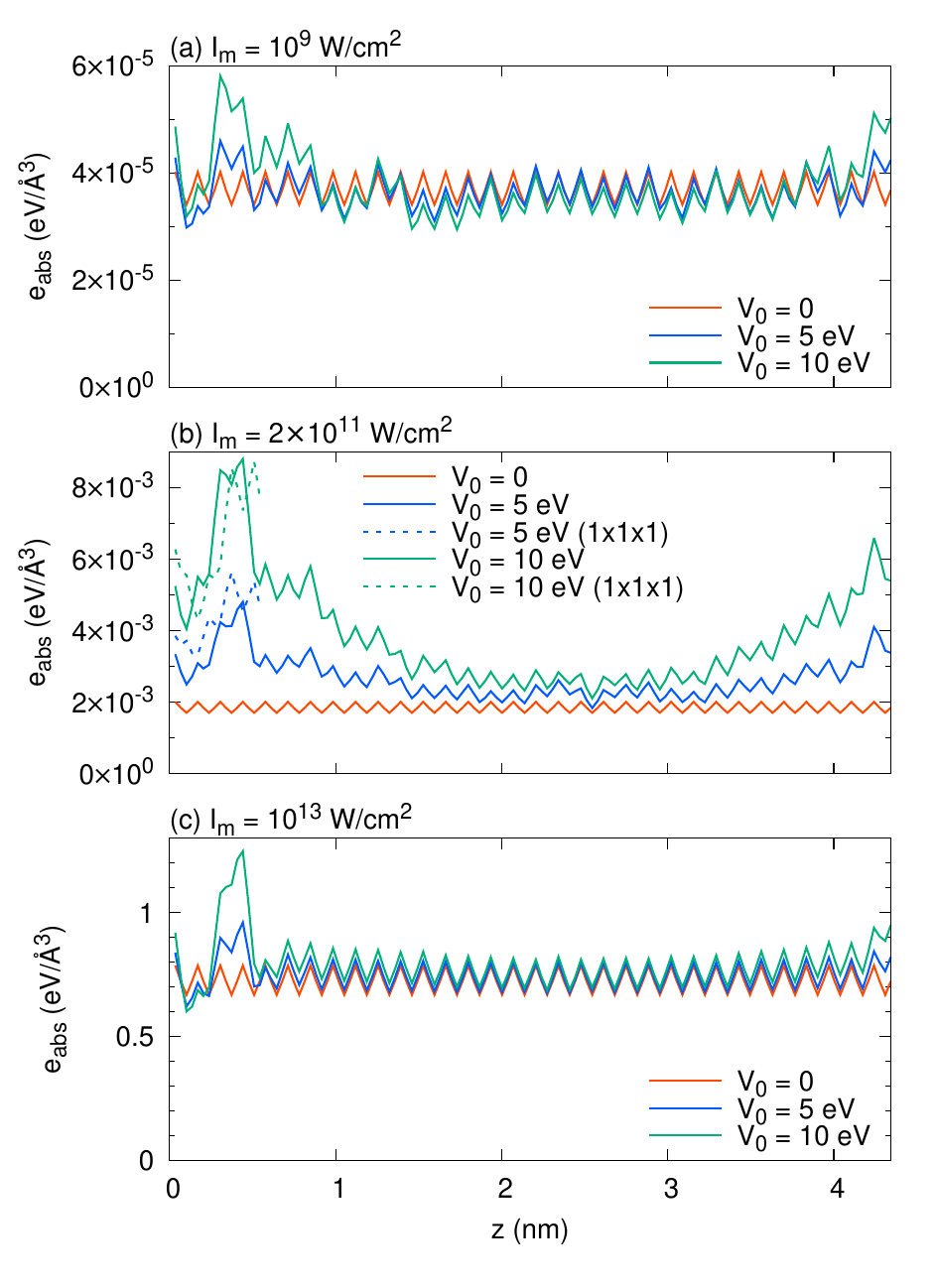}
    \caption{\label{fig:ext_eabs}
    (a--c) Absorbed energy density for intensities of $I_m=2\times 10^{11}$, $10^9$, and $10^{13}$ W/cm$^2$, respectively, at $T_e=1$ eV under the static external potential.
    }
\end{figure}

Finally, we investigate the causes of the surface-induced enhancement of the absorption.
At a solid surface, there is a step-wise potential change at the boundary of the vacuum and material.
For the Si surface, the difference in the Kohn-Sham local potential between the bulk region and vacuum is about 20 eV.
This local potential change at the surface may affect the absorbed energy density.

To consider that effect, we shall introduce the following static external potential:
\begin{equation}
    V_{\rm ext}(z)=
  \begin{cases}
    V_0 \sin(2\pi z/a), & (0 < z <a), \\
 0, & ({\rm otherwise}),
  \end{cases}
  \label{eq:V_ext}
\end{equation}
where $V_0$ is a parameter for the potential strength and $a=0.543$ nm is the lattice constant of Si.
If we apply $V_{\rm ext}(z)$ to the Si bulk locally, a similar effect of the potential change can be artificially reproduced in a region around $0 < z <a$.
In other words, the surface effect is mimicked by using $V_{\rm ext}(z)$ in the bulk region artificially.
We employ the potential form of Eq.~(\ref{eq:V_ext}) because it is a simple continuous function with an average value of zero to prevent electron accumulation.
Here, we consider a supercell of $1\times 1\times 8$ bulk Si under the external potential $V_{\rm ext}(z)$ at the temperature $T_e=1$ eV.
The $z$ coordinate for the supercell is in $[0,8a]$ and $V_{\rm ext}(z)$ has a nonzero value within $[0,a]$ only. 
The ground state orbitals $\{u^{\rm GS}_{n{\bf k}}\}$ are solved in the supercell under the presence of $V_{\rm ext}(z)$.
The TDKS equation Eq.~(\ref{eq:tdks}) is solved with the one-electron Hamiltonian including $V_{\rm ext}(z)$.

Figures~\ref{fig:ext_lr}(a,b) provide the $xx$ component of the dielectric function at $T_e=1$ eV with several strengths of $V_0=0$, 5, and 10 eV.
Here, the case of $V_0=0$ corresponds to the ordinary bulk Si.
The dielectric function remains almost unchanged by $V_{\rm ext}(z)$, except that the peak of the imaginary part around 4 eV is slightly reduced.
This behavior is similar to that of the slab dielectric function in Figs.~\ref{fig:lr}(c,d), where the peak of the slab [Fig.~\ref{fig:lr}(d)] is slightly lower than the corresponding peak of the bulk [Fig.~\ref{fig:lr}(b)] at each temperature.
The potential changes in the $z$ axis (vertical direction) may have little impact on the linear response along the $x$ axis (horizontal direction) because the symmetry in the $xy$ plane is not affected.

Figure~\ref{fig:ext_lr}(c) is the same as Fig.~\ref{fig:slab}(a) but for the supercell with $V_{\rm ext}(z)$.
As in the comparison between the slab and bulk, the three $V_0$ cases coincide with each other below $I_m=10^{10}$ W/cm$^2$ and above $5\times 10^{12}$ W/cm$^2$.
The excitation energy increases monotonically as $V_0$ increases around the FTSA intensity $I_m=2\times 10^{11}$ W/cm$^2$.
The finite $V_0$ cases have similar properties to the slab excitation energy.

Figure~\ref{fig:ext_eabs} shows the final absorbed energy density $e_{\rm abs}(z,t=\tau_p)$ at the three intensities under the same conditions as in Fig.~\ref{fig:slab_eabs}(b) but for the supercell with $V_{\rm ext}(z)$.
The ordinary bulk results are plotted as a red solid line, which exhibits fine oscillations of the Si-Si bond as in Fig.~\ref{fig:slab_eabs}.
In the low-intensity case [Fig.~\ref{fig:ext_eabs}(a)], as expected from the dielectric function, $V_{\rm ext}(z)$ has little impact on the absorption but slight deviations are located around $[0,a]$.
For the FTSA intensity [Fig.~\ref{fig:ext_eabs}(b)], the deviations become apparent in an extended region.
The absorbed energy densities around $[0,a]$ in the finite $V_0$ cases are several times larger than that in the ordinary bulk case of $V_0=0$.
The tail of the deviations appears comparable to those observed at the slab surfaces in the FTSA cases such as Fig.~\ref{fig:slab_eabs}(a).
In Fig.~\ref{fig:ext_eabs}(b), results for the usual $1\times 1\times 1$ unit cell with $V_{\rm ext}(z)$ are plotted as dashed lines.
As these results roughly agree with the corresponding supercell cases, it is confirmed that the effect of the potential change is local. 
In the very high-intensity case [Fig.~\ref{fig:ext_eabs}(c)], deviations become small again.
The overall behavior in Fig.~\ref{fig:ext_eabs} is very similar to that in Fig.~\ref{fig:slab_eabs}(b).

From the above results, the surface-induced enhancement in the slab cases is consistent with the enhancement mechanism by the local potential change.  
In the slab cases, the peaks of the absorbed energy density are located at a depth of around 1 nm from the surfaces rather than at the surfaces.
This may be because the electron density is almost absent at the surface unlike the supercell with $V_{\rm ext}(z)$, where there is no surface.
For both the slab and supercell cases, the absorption coincides with that for the usual bulk case in the linear regime where the one-photon absorption is dominant and in the highly nonlinear regime where the two-photon absorption is dominant.
The absorption enhancement, or suppression of FTSA, by local potential changes occurs only under the FTSA conditions.
As FTSA stems from the Pauli blocking in the respective conduction and valence bands (Sec.~\ref{sec:toy}), it is suggested that the local potential change alters conditions of the Pauli blocking by breaking the inversion symmetry of the wavefunctions.

\section{Conclusion \label{sec:conclusion}}

Using the TDDFT-based first-principles methods, we systematically investigated the electronic excitation in Si under the presence of a near-infrared femtosecond laser pulse at finite electron temperatures.
In the bulk region, we found that the saturable absorption occurs at a certain laser intensity, but it does not occur at zero temperature.
We referred to it as the finite-temperature saturable absorption (FTSA).
From the toy model analysis, we confirmed that FTSA is explained by the Pauli blocking of one-photon absorption corresponding to conduction-to-conduction and valence-to-valence transitions.
With higher intensities, the multi-photon excitation across the bandgap overwhelms the one-photon excitation and FTSA disappears.

By comparing the bulk and slab results, we showed that the Si surface suppresses FTSA and enhances the absorbed energy near the surface from that in the bulk region.
This effect is simulated in the bulk region using the static external potential that mimics the surface potential change.
We concluded that the FTSA suppression at the surface is caused by the symmetry breaking of the wavefunctions by the local potential change.
The migration of the electronic energy density from the surface region to the bulk region is also elucidated with the frozen Hamiltonian formalism.

We discussed nonlinear light propagation in finite-temperature Si slabs using the Maxwell-TDDFT methods, which combine the Maxwell equations for the electromagnetic field with TDDFT for electrons.
We confirmed that the light propagation effect can be considered macroscopically in the spatial scale of the laser wavelength and separated from the microscopic electron dynamics.
The surface effect can be separately evaluated with the microscopic slab calculations.

Our findings reveal a cumulative effect of laser irradiation on solid surfaces that has not been previously recognized.
In widely used numerical models for laser ablation such as the two-temperature model, the nonlinear light absorption is assumed to be the sum of single- and multi-photon processes. 
FTSA cannot be taken into account in such models by straightforward extension because it is a quantum effect relating to electron occupancy. 
The present study suggests that existing numerical models may fail to predict the absorption of laser pulses under certain conditions.


\begin{acknowledgements}
The authors thank Dr.~Thanh Hung Dinh (QST), Dr.~Masayasu Hata (QST), Dr.~Mizuki Tani (QST), Prof.~Mitsuharu Uemoto (Kobe University), and Prof.~Kazuhiro Yabana (University of Tsukuba) for fruitful discussions.
We acknowledge the critical reading of the manuscript by Dr.~Ryuji Itakura (QST).
This research was supported by MEXT Quantum Leap Flagship Program (MEXT Q-LEAP) Grant Number JPMXS0118067246, and by JSPS KAKENHI Grant Number 24K01224 and 24K17629. 
Calculations are carried out on Fugaku supercomputer under the support through the HPCI System Research Project (Project ID: hp220120 and hp230273), SGI8600 at Japan Atomic Energy Agency (JAEA), and Wisteria at the University of Tokyo under the support by Multidisciplinary Cooperative Research Program in CCS, University of Tsukuba.
\end{acknowledgements}


%


\begin{thebibliography}{26}%
\makeatletter
\providecommand \@ifxundefined [1]{%
 \@ifx{#1\undefined}
}%
\providecommand \@ifnum [1]{%
 \ifnum #1\expandafter \@firstoftwo
 \else \expandafter \@secondoftwo
 \fi
}%
\providecommand \@ifx [1]{%
 \ifx #1\expandafter \@firstoftwo
 \else \expandafter \@secondoftwo
 \fi
}%
\providecommand \natexlab [1]{#1}%
\providecommand \enquote  [1]{``#1''}%
\providecommand \bibnamefont  [1]{#1}%
\providecommand \bibfnamefont [1]{#1}%
\providecommand \citenamefont [1]{#1}%
\providecommand \href@noop [0]{\@secondoftwo}%
\providecommand \href [0]{\begingroup \@sanitize@url \@href}%
\providecommand \@href[1]{\@@startlink{#1}\@@href}%
\providecommand \@@href[1]{\endgroup#1\@@endlink}%
\providecommand \@sanitize@url [0]{\catcode `\\12\catcode `\$12\catcode `\&12\catcode `\#12\catcode `\^12\catcode `\_12\catcode `\%12\relax}%
\providecommand \@@startlink[1]{}%
\providecommand \@@endlink[0]{}%
\providecommand \url  [0]{\begingroup\@sanitize@url \@url }%
\providecommand \@url [1]{\endgroup\@href {#1}{\urlprefix }}%
\providecommand \urlprefix  [0]{URL }%
\providecommand \Eprint [0]{\href }%
\providecommand \doibase [0]{http://dx.doi.org/}%
\providecommand \selectlanguage [0]{\@gobble}%
\providecommand \bibinfo  [0]{\@secondoftwo}%
\providecommand \bibfield  [0]{\@secondoftwo}%
\providecommand \translation [1]{[#1]}%
\providecommand \BibitemOpen [0]{}%
\providecommand \bibitemStop [0]{}%
\providecommand \bibitemNoStop [0]{.\EOS\space}%
\providecommand \EOS [0]{\spacefactor3000\relax}%
\providecommand \BibitemShut  [1]{\csname bibitem#1\endcsname}%
\let\auto@bib@innerbib\@empty
\bibitem [{\citenamefont {Tien}\ \emph {et~al.}(1999)\citenamefont {Tien}, \citenamefont {Backus}, \citenamefont {Kapteyn}, \citenamefont {Murnane},\ and\ \citenamefont {Mourou}}]{Tien1999}%
  \BibitemOpen
  \bibfield  {author} {\bibinfo {author} {\bibfnamefont {A.-C.}\ \bibnamefont {Tien}}, \bibinfo {author} {\bibfnamefont {S.}~\bibnamefont {Backus}}, \bibinfo {author} {\bibfnamefont {H.}~\bibnamefont {Kapteyn}}, \bibinfo {author} {\bibfnamefont {M.}~\bibnamefont {Murnane}}, \ and\ \bibinfo {author} {\bibfnamefont {G.}~\bibnamefont {Mourou}},\ }\href {\doibase 10.1103/PhysRevLett.82.3883} {\bibfield  {journal} {\bibinfo  {journal} {Physical Review Letters}\ }\textbf {\bibinfo {volume} {82}},\ \bibinfo {pages} {3883} (\bibinfo {year} {1999})}\BibitemShut {NoStop}%
\bibitem [{\citenamefont {Bonse}\ \emph {et~al.}(2002)\citenamefont {Bonse}, \citenamefont {Baudach}, \citenamefont {Kr{\"{u}}ger}, \citenamefont {Kautek},\ and\ \citenamefont {Lenzner}}]{Bonse2002}%
  \BibitemOpen
  \bibfield  {author} {\bibinfo {author} {\bibfnamefont {J.}~\bibnamefont {Bonse}}, \bibinfo {author} {\bibfnamefont {S.}~\bibnamefont {Baudach}}, \bibinfo {author} {\bibfnamefont {J.}~\bibnamefont {Kr{\"{u}}ger}}, \bibinfo {author} {\bibfnamefont {W.}~\bibnamefont {Kautek}}, \ and\ \bibinfo {author} {\bibfnamefont {M.}~\bibnamefont {Lenzner}},\ }\href {\doibase 10.1007/s003390100893} {\bibfield  {journal} {\bibinfo  {journal} {Appl. Phys. A}\ }\textbf {\bibinfo {volume} {74}},\ \bibinfo {pages} {19} (\bibinfo {year} {2002})}\BibitemShut {NoStop}%
\bibitem [{\citenamefont {Gattass}\ and\ \citenamefont {Mazur}(2008)}]{Gattass2008}%
  \BibitemOpen
  \bibfield  {author} {\bibinfo {author} {\bibfnamefont {R.~R.}\ \bibnamefont {Gattass}}\ and\ \bibinfo {author} {\bibfnamefont {E.}~\bibnamefont {Mazur}},\ }\href {\doibase 10.1038/nphoton.2008.47} {\bibfield  {journal} {\bibinfo  {journal} {Nature Photonics}\ }\textbf {\bibinfo {volume} {2}},\ \bibinfo {pages} {219} (\bibinfo {year} {2008})}\BibitemShut {NoStop}%
\bibitem [{\citenamefont {Kerse}\ \emph {et~al.}(2016)\citenamefont {Kerse}, \citenamefont {Kalayc^^c4^^b1o^^c4^^9flu}, \citenamefont {Elahi}, \citenamefont {^^c3^^87etin}, \citenamefont {Kesim}, \citenamefont {^^c3^^96nder Ak^^c3^^a7aalan}, \citenamefont {Yava^^c5^^9f}, \citenamefont {A^^c5^^9f^^c4^^b1k}, \citenamefont {^^c3^^96ktem}, \citenamefont {Hoogland}, \citenamefont {Holzwarth},\ and\ \citenamefont {^^c3^^96mer Ilday}}]{Kerse2016}%
  \BibitemOpen
  \bibfield  {author} {\bibinfo {author} {\bibfnamefont {C.}~\bibnamefont {Kerse}}, \bibinfo {author} {\bibfnamefont {H.}~\bibnamefont {Kalayc^^c4^^b1o^^c4^^9flu}}, \bibinfo {author} {\bibfnamefont {P.}~\bibnamefont {Elahi}}, \bibinfo {author} {\bibfnamefont {B.}~\bibnamefont {^^c3^^87etin}}, \bibinfo {author} {\bibfnamefont {D.~K.}\ \bibnamefont {Kesim}}, \bibinfo {author} {\bibnamefont {^^c3^^96nder Ak^^c3^^a7aalan}}, \bibinfo {author} {\bibfnamefont {S.}~\bibnamefont {Yava^^c5^^9f}}, \bibinfo {author} {\bibfnamefont {M.~D.}\ \bibnamefont {A^^c5^^9f^^c4^^b1k}}, \bibinfo {author} {\bibfnamefont {B.}~\bibnamefont {^^c3^^96ktem}}, \bibinfo {author} {\bibfnamefont {H.}~\bibnamefont {Hoogland}}, \bibinfo {author} {\bibfnamefont {R.}~\bibnamefont {Holzwarth}}, \ and\ \bibinfo {author} {\bibfnamefont {F.}~\bibnamefont {^^c3^^96mer Ilday}},\ }\href {\doibase 10.1038/nature18619} {\bibfield  {journal} {\bibinfo  {journal} {Nature}\ }\textbf {\bibinfo {volume} {537}},\ \bibinfo {pages} {84} (\bibinfo {year}
  {2016})}\BibitemShut {NoStop}%
\bibitem [{\citenamefont {Obata}\ \emph {et~al.}(2023)\citenamefont {Obata}, \citenamefont {Caballero-Lucas}, \citenamefont {Kawabata}, \citenamefont {Miyaji},\ and\ \citenamefont {Sugioka}}]{Obata2023}%
  \BibitemOpen
  \bibfield  {author} {\bibinfo {author} {\bibfnamefont {K.}~\bibnamefont {Obata}}, \bibinfo {author} {\bibfnamefont {F.}~\bibnamefont {Caballero-Lucas}}, \bibinfo {author} {\bibfnamefont {S.}~\bibnamefont {Kawabata}}, \bibinfo {author} {\bibfnamefont {G.}~\bibnamefont {Miyaji}}, \ and\ \bibinfo {author} {\bibfnamefont {K.}~\bibnamefont {Sugioka}},\ }\href {\doibase 10.1088/2631-7990/acc0e5} {\bibfield  {journal} {\bibinfo  {journal} {International Journal of Extreme Manufacturing}\ }\textbf {\bibinfo {volume} {5}},\ \bibinfo {pages} {025002} (\bibinfo {year} {2023})}\BibitemShut {NoStop}%
\bibitem [{\citenamefont {Goldman}\ and\ \citenamefont {Prybyla}(1994)}]{Goldman1994}%
  \BibitemOpen
  \bibfield  {author} {\bibinfo {author} {\bibfnamefont {J.~R.}\ \bibnamefont {Goldman}}\ and\ \bibinfo {author} {\bibfnamefont {J.~A.}\ \bibnamefont {Prybyla}},\ }\href {\doibase 10.1103/PhysRevLett.72.1364} {\bibfield  {journal} {\bibinfo  {journal} {Physical Review Letters}\ }\textbf {\bibinfo {volume} {72}},\ \bibinfo {pages} {1364} (\bibinfo {year} {1994})}\BibitemShut {NoStop}%
\bibitem [{\citenamefont {Mueller}\ and\ \citenamefont {Rethfeld}(2013)}]{Mueller2013}%
  \BibitemOpen
  \bibfield  {author} {\bibinfo {author} {\bibfnamefont {B.~Y.}\ \bibnamefont {Mueller}}\ and\ \bibinfo {author} {\bibfnamefont {B.}~\bibnamefont {Rethfeld}},\ }\href {\doibase 10.1103/PhysRevB.87.035139} {\bibfield  {journal} {\bibinfo  {journal} {Physical Review B}\ }\textbf {\bibinfo {volume} {87}},\ \bibinfo {pages} {035139} (\bibinfo {year} {2013})}\BibitemShut {NoStop}%
\bibitem [{\citenamefont {R^^c3^^a4mer}\ \emph {et~al.}(2014)\citenamefont {R^^c3^^a4mer}, \citenamefont {Osmani},\ and\ \citenamefont {Rethfeld}}]{Ramer2014}%
  \BibitemOpen
  \bibfield  {author} {\bibinfo {author} {\bibfnamefont {A.}~\bibnamefont {R^^c3^^a4mer}}, \bibinfo {author} {\bibfnamefont {O.}~\bibnamefont {Osmani}}, \ and\ \bibinfo {author} {\bibfnamefont {B.}~\bibnamefont {Rethfeld}},\ }\href {\doibase 10.1063/1.4891633} {\bibfield  {journal} {\bibinfo  {journal} {Journal of Applied Physics}\ }\textbf {\bibinfo {volume} {116}},\ \bibinfo {pages} {053508} (\bibinfo {year} {2014})}\BibitemShut {NoStop}%
\bibitem [{\citenamefont {Rethfeld}\ \emph {et~al.}(2017)\citenamefont {Rethfeld}, \citenamefont {Ivanov}, \citenamefont {Garcia},\ and\ \citenamefont {Anisimov}}]{Rethfeld2017}%
  \BibitemOpen
  \bibfield  {author} {\bibinfo {author} {\bibfnamefont {B.}~\bibnamefont {Rethfeld}}, \bibinfo {author} {\bibfnamefont {D.~S.}\ \bibnamefont {Ivanov}}, \bibinfo {author} {\bibfnamefont {M.~E.}\ \bibnamefont {Garcia}}, \ and\ \bibinfo {author} {\bibfnamefont {S.~I.}\ \bibnamefont {Anisimov}},\ }\href {\doibase 10.1088/1361-6463/50/19/193001} {\bibfield  {journal} {\bibinfo  {journal} {Journal of Physics D: Applied Physics}\ }\textbf {\bibinfo {volume} {50}},\ \bibinfo {pages} {193001} (\bibinfo {year} {2017})}\BibitemShut {NoStop}%
\bibitem [{\citenamefont {Sato}\ \emph {et~al.}(2014)\citenamefont {Sato}, \citenamefont {Shinohara}, \citenamefont {Otobe},\ and\ \citenamefont {Yabana}}]{Sato2014}%
  \BibitemOpen
  \bibfield  {author} {\bibinfo {author} {\bibfnamefont {S.~A.}\ \bibnamefont {Sato}}, \bibinfo {author} {\bibfnamefont {Y.}~\bibnamefont {Shinohara}}, \bibinfo {author} {\bibfnamefont {T.}~\bibnamefont {Otobe}}, \ and\ \bibinfo {author} {\bibfnamefont {K.}~\bibnamefont {Yabana}},\ }\href {\doibase 10.1103/PhysRevB.90.174303} {\bibfield  {journal} {\bibinfo  {journal} {Physical Review B}\ }\textbf {\bibinfo {volume} {90}},\ \bibinfo {pages} {174303} (\bibinfo {year} {2014})}\BibitemShut {NoStop}%
\bibitem [{\citenamefont {Runge}\ and\ \citenamefont {Gross}(1984)}]{Runge1984}%
  \BibitemOpen
  \bibfield  {author} {\bibinfo {author} {\bibfnamefont {E.}~\bibnamefont {Runge}}\ and\ \bibinfo {author} {\bibfnamefont {E.~K.~U.}\ \bibnamefont {Gross}},\ }\href {\doibase 10.1103/PhysRevLett.52.997} {\bibfield  {journal} {\bibinfo  {journal} {Phys. Rev. Lett.}\ }\textbf {\bibinfo {volume} {52}},\ \bibinfo {pages} {997} (\bibinfo {year} {1984})}\BibitemShut {NoStop}%
\bibitem [{\citenamefont {Yabana}\ \emph {et~al.}(2012)\citenamefont {Yabana}, \citenamefont {Sugiyama}, \citenamefont {Shinohara}, \citenamefont {Otobe},\ and\ \citenamefont {Bertsch}}]{Yabana2012}%
  \BibitemOpen
  \bibfield  {author} {\bibinfo {author} {\bibfnamefont {K.}~\bibnamefont {Yabana}}, \bibinfo {author} {\bibfnamefont {T.}~\bibnamefont {Sugiyama}}, \bibinfo {author} {\bibfnamefont {Y.}~\bibnamefont {Shinohara}}, \bibinfo {author} {\bibfnamefont {T.}~\bibnamefont {Otobe}}, \ and\ \bibinfo {author} {\bibfnamefont {G.~F.}\ \bibnamefont {Bertsch}},\ }\href {\doibase 10.1103/PhysRevB.85.045134} {\bibfield  {journal} {\bibinfo  {journal} {Phys. Rev. B}\ }\textbf {\bibinfo {volume} {85}},\ \bibinfo {pages} {045134} (\bibinfo {year} {2012})}\BibitemShut {NoStop}%
\bibitem [{\citenamefont {Yamada}\ \emph {et~al.}(2018)\citenamefont {Yamada}, \citenamefont {Noda}, \citenamefont {Nobusada},\ and\ \citenamefont {Yabana}}]{Yamada2018}%
  \BibitemOpen
  \bibfield  {author} {\bibinfo {author} {\bibfnamefont {S.}~\bibnamefont {Yamada}}, \bibinfo {author} {\bibfnamefont {M.}~\bibnamefont {Noda}}, \bibinfo {author} {\bibfnamefont {K.}~\bibnamefont {Nobusada}}, \ and\ \bibinfo {author} {\bibfnamefont {K.}~\bibnamefont {Yabana}},\ }\href {\doibase 10.1103/PhysRevB.98.245147} {\bibfield  {journal} {\bibinfo  {journal} {Phys. Rev. B}\ }\textbf {\bibinfo {volume} {98}},\ \bibinfo {pages} {245147} (\bibinfo {year} {2018})}\BibitemShut {NoStop}%
\bibitem [{\citenamefont {Bertsch}\ \emph {et~al.}(2000)\citenamefont {Bertsch}, \citenamefont {Iwata}, \citenamefont {Rubio},\ and\ \citenamefont {Yabana}}]{Bertsch2000}%
  \BibitemOpen
  \bibfield  {author} {\bibinfo {author} {\bibfnamefont {G.~F.}\ \bibnamefont {Bertsch}}, \bibinfo {author} {\bibfnamefont {J.-I.}\ \bibnamefont {Iwata}}, \bibinfo {author} {\bibfnamefont {A.}~\bibnamefont {Rubio}}, \ and\ \bibinfo {author} {\bibfnamefont {K.}~\bibnamefont {Yabana}},\ }\href {\doibase 10.1103/PhysRevB.62.7998} {\bibfield  {journal} {\bibinfo  {journal} {Phys. Rev. B}\ }\textbf {\bibinfo {volume} {62}},\ \bibinfo {pages} {7998} (\bibinfo {year} {2000})}\BibitemShut {NoStop}%
\bibitem [{\citenamefont {Otobe}\ \emph {et~al.}(2008)\citenamefont {Otobe}, \citenamefont {Yamagiwa}, \citenamefont {Iwata}, \citenamefont {Yabana}, \citenamefont {Nakatsukasa},\ and\ \citenamefont {Bertsch}}]{Otobe2008}%
  \BibitemOpen
  \bibfield  {author} {\bibinfo {author} {\bibfnamefont {T.}~\bibnamefont {Otobe}}, \bibinfo {author} {\bibfnamefont {M.}~\bibnamefont {Yamagiwa}}, \bibinfo {author} {\bibfnamefont {J.-I.}\ \bibnamefont {Iwata}}, \bibinfo {author} {\bibfnamefont {K.}~\bibnamefont {Yabana}}, \bibinfo {author} {\bibfnamefont {T.}~\bibnamefont {Nakatsukasa}}, \ and\ \bibinfo {author} {\bibfnamefont {G.~F.}\ \bibnamefont {Bertsch}},\ }\href {\doibase 10.1103/PhysRevB.77.165104} {\bibfield  {journal} {\bibinfo  {journal} {Phys. Rev. B}\ }\textbf {\bibinfo {volume} {77}},\ \bibinfo {pages} {165104} (\bibinfo {year} {2008})}\BibitemShut {NoStop}%
\bibitem [{\citenamefont {Troullier}\ and\ \citenamefont {Martins}(1991)}]{Troullier1991}%
  \BibitemOpen
  \bibfield  {author} {\bibinfo {author} {\bibfnamefont {N.}~\bibnamefont {Troullier}}\ and\ \bibinfo {author} {\bibfnamefont {J.~L.}\ \bibnamefont {Martins}},\ }\href@noop {} {\bibfield  {journal} {\bibinfo  {journal} {Physical review B}\ }\textbf {\bibinfo {volume} {43}},\ \bibinfo {pages} {1993} (\bibinfo {year} {1991})}\BibitemShut {NoStop}%
\bibitem [{\citenamefont {Kleinman}\ and\ \citenamefont {Bylander}(1982)}]{Kleinman1982}%
  \BibitemOpen
  \bibfield  {author} {\bibinfo {author} {\bibfnamefont {L.}~\bibnamefont {Kleinman}}\ and\ \bibinfo {author} {\bibfnamefont {D.~M.}\ \bibnamefont {Bylander}},\ }\href {\doibase 10.1103/PhysRevLett.48.1425} {\bibfield  {journal} {\bibinfo  {journal} {Phys. Rev. Lett.}\ }\textbf {\bibinfo {volume} {48}},\ \bibinfo {pages} {1425} (\bibinfo {year} {1982})}\BibitemShut {NoStop}%
\bibitem [{\citenamefont {Perdew}\ and\ \citenamefont {Zunger}(1981)}]{Perdew1981}%
  \BibitemOpen
  \bibfield  {author} {\bibinfo {author} {\bibfnamefont {J.~P.}\ \bibnamefont {Perdew}}\ and\ \bibinfo {author} {\bibfnamefont {A.}~\bibnamefont {Zunger}},\ }\href@noop {} {\bibfield  {journal} {\bibinfo  {journal} {Physical Review B}\ }\textbf {\bibinfo {volume} {23}},\ \bibinfo {pages} {5048} (\bibinfo {year} {1981})}\BibitemShut {NoStop}%
\bibitem [{\citenamefont {Yamada}\ and\ \citenamefont {Yabana}(2021)}]{Yamada2021}%
  \BibitemOpen
  \bibfield  {author} {\bibinfo {author} {\bibfnamefont {S.}~\bibnamefont {Yamada}}\ and\ \bibinfo {author} {\bibfnamefont {K.}~\bibnamefont {Yabana}},\ }\href {\doibase 10.1103/PhysRevB.103.155426} {\bibfield  {journal} {\bibinfo  {journal} {Phys. Rev. B}\ }\textbf {\bibinfo {volume} {103}},\ \bibinfo {pages} {155426} (\bibinfo {year} {2021})}\BibitemShut {NoStop}%
\bibitem [{\citenamefont {Sato}(2023)}]{Sato2023}%
  \BibitemOpen
  \bibfield  {author} {\bibinfo {author} {\bibfnamefont {S.~A.}\ \bibnamefont {Sato}},\ }\href {\doibase 10.7566/JPSJ.92.094401} {\bibfield  {journal} {\bibinfo  {journal} {Journal of the Physical Society of Japan}\ }\textbf {\bibinfo {volume} {92}},\ \bibinfo {pages} {094401} (\bibinfo {year} {2023})}\BibitemShut {NoStop}%
\bibitem [{\citenamefont {Noda}\ \emph {et~al.}(2019)\citenamefont {Noda}, \citenamefont {Sato}, \citenamefont {Hirokawa}, \citenamefont {Uemoto}, \citenamefont {Takeuchi}, \citenamefont {Yamada}, \citenamefont {Yamada}, \citenamefont {Shinohara}, \citenamefont {Yamaguchi}, \citenamefont {Iida}, \citenamefont {Floss}, \citenamefont {Otobe}, \citenamefont {Lee}, \citenamefont {Ishimura}, \citenamefont {Boku}, \citenamefont {Bertsch}, \citenamefont {Nobusada},\ and\ \citenamefont {Yabana}}]{Noda2019}%
  \BibitemOpen
  \bibfield  {author} {\bibinfo {author} {\bibfnamefont {M.}~\bibnamefont {Noda}}, \bibinfo {author} {\bibfnamefont {S.~A.}\ \bibnamefont {Sato}}, \bibinfo {author} {\bibfnamefont {Y.}~\bibnamefont {Hirokawa}}, \bibinfo {author} {\bibfnamefont {M.}~\bibnamefont {Uemoto}}, \bibinfo {author} {\bibfnamefont {T.}~\bibnamefont {Takeuchi}}, \bibinfo {author} {\bibfnamefont {S.}~\bibnamefont {Yamada}}, \bibinfo {author} {\bibfnamefont {A.}~\bibnamefont {Yamada}}, \bibinfo {author} {\bibfnamefont {Y.}~\bibnamefont {Shinohara}}, \bibinfo {author} {\bibfnamefont {M.}~\bibnamefont {Yamaguchi}}, \bibinfo {author} {\bibfnamefont {K.}~\bibnamefont {Iida}}, \bibinfo {author} {\bibfnamefont {I.}~\bibnamefont {Floss}}, \bibinfo {author} {\bibfnamefont {T.}~\bibnamefont {Otobe}}, \bibinfo {author} {\bibfnamefont {K.-M. K.-M.}\ \bibnamefont {Lee}}, \bibinfo {author} {\bibfnamefont {K.}~\bibnamefont {Ishimura}}, \bibinfo {author} {\bibfnamefont {T.}~\bibnamefont {Boku}}, \bibinfo {author} {\bibfnamefont {G.~F.}\ \bibnamefont
  {Bertsch}}, \bibinfo {author} {\bibfnamefont {K.}~\bibnamefont {Nobusada}}, \ and\ \bibinfo {author} {\bibfnamefont {K.}~\bibnamefont {Yabana}},\ }\href {\doibase 10.1016/j.cpc.2018.09.018} {\bibfield  {journal} {\bibinfo  {journal} {Comput. Phys. Commun.}\ }\textbf {\bibinfo {volume} {235}},\ \bibinfo {pages} {356} (\bibinfo {year} {2019})}\BibitemShut {NoStop}%
\bibitem [{SAL()}]{SALMON_web}%
  \BibitemOpen
  \href@noop {} {}\bibinfo {howpublished} {\url{https://salmon-tddft.jp}},\ \bibinfo {note} {{SALMON} official website}\BibitemShut {NoStop}%
\bibitem [{\citenamefont {Yabana}\ and\ \citenamefont {Bertsch}(1996)}]{Yabana1996}%
  \BibitemOpen
  \bibfield  {author} {\bibinfo {author} {\bibfnamefont {K.}~\bibnamefont {Yabana}}\ and\ \bibinfo {author} {\bibfnamefont {G.~F.}\ \bibnamefont {Bertsch}},\ }\href {\doibase 10.1103/PhysRevB.54.4484} {\bibfield  {journal} {\bibinfo  {journal} {Phys. Rev. B}\ }\textbf {\bibinfo {volume} {54}},\ \bibinfo {pages} {4484} (\bibinfo {year} {1996})}\BibitemShut {NoStop}%
\bibitem [{\citenamefont {Yamada}\ and\ \citenamefont {Yabana}(2019)}]{Yamada2019}%
  \BibitemOpen
  \bibfield  {author} {\bibinfo {author} {\bibfnamefont {A.}~\bibnamefont {Yamada}}\ and\ \bibinfo {author} {\bibfnamefont {K.}~\bibnamefont {Yabana}},\ }\href {\doibase 10.1140/epjd/e2019-90334-7} {\bibfield  {journal} {\bibinfo  {journal} {The European Physical Journal D}\ }\textbf {\bibinfo {volume} {73}},\ \bibinfo {pages} {87} (\bibinfo {year} {2019})}\BibitemShut {NoStop}%
\bibitem [{\citenamefont {Uemoto}\ \emph {et~al.}(2021)\citenamefont {Uemoto}, \citenamefont {Kurata}, \citenamefont {Kawaguchi},\ and\ \citenamefont {Yabana}}]{Uemoto2021}%
  \BibitemOpen
  \bibfield  {author} {\bibinfo {author} {\bibfnamefont {M.}~\bibnamefont {Uemoto}}, \bibinfo {author} {\bibfnamefont {S.}~\bibnamefont {Kurata}}, \bibinfo {author} {\bibfnamefont {N.}~\bibnamefont {Kawaguchi}}, \ and\ \bibinfo {author} {\bibfnamefont {K.}~\bibnamefont {Yabana}},\ }\href {\doibase 10.1103/PhysRevB.103.085433} {\bibfield  {journal} {\bibinfo  {journal} {Phys. Rev. B}\ }\textbf {\bibinfo {volume} {103}},\ \bibinfo {pages} {085433} (\bibinfo {year} {2021})}\BibitemShut {NoStop}%
\bibitem [{\citenamefont {Yamada}\ and\ \citenamefont {Yabana}(2024)}]{Yamada2024}%
  \BibitemOpen
  \bibfield  {author} {\bibinfo {author} {\bibfnamefont {A.}~\bibnamefont {Yamada}}\ and\ \bibinfo {author} {\bibfnamefont {K.}~\bibnamefont {Yabana}},\ }\href {\doibase 10.1103/PhysRevB.109.245130} {\bibfield  {journal} {\bibinfo  {journal} {Physical Review B}\ }\textbf {\bibinfo {volume} {109}},\ \bibinfo {pages} {245130} (\bibinfo {year} {2024})}\BibitemShut {NoStop}%
\end{thebibliography}

\end{document}